# Room temperature spin Hall effect in graphene/MoS$_2$ van der Waals heterostructures

C. K. Safeer[1,‡], Josep Ingla-Aynés[1,‡], Franz Herling[1,‡], José H. Garcia[2], Marc Vila[2,3], Nerea Ontoso[1], M. Reyes Calvo[1,4], Stephan Roche[2,5], Luis E. Hueso[1,4], Fèlix Casanova[1,4,*].

[1] CIC nanoGUNE, 20018 Donostia-San Sebastian, Basque Country, Spain.
[2] Catalan Institute of Nanoscience and Nanotechnology (ICN2), CSIC and The Barcelona Institute of Science and Technology, Campus UAB, 08193 Bellaterra, Catalonia, Spain.
[3] Department of Physics, Universitat Autònoma de Barcelona, Campus UAB, 08193 Bellaterra, Catalonia, Spain.
[4] IKERBASQUE, Basque Foundation for Science, 48013 Bilbao, Basque Country, Spain.
[5] ICREA – Institució Catalana de Recerca i Estudis Avançats, 08010 Barcelona, Catalonia, Spain.

[‡]These authors contributed equally to this work
*E-mail: f.casanova@nanogune.eu

**Keywords:** Graphene, Transition metal dichalcogenides, Spintronics, Spin Hall effect, Rashba-Edelstein effect.

**Graphene is an excellent material for long distance spin transport but allows little spin manipulation. Transition metal dichalcogenides imprint their strong spin-orbit coupling into graphene via proximity effect, and it has been predicted that efficient spin-to-charge conversion due to spin Hall and Rashba-Edelstein effects could be achieved. Here, by combining Hall probes with ferromagnetic electrodes, we unambiguously demonstrate experimentally spin Hall effect in graphene induced by MoS$_2$ proximity and for varying temperature up to room temperature. The fact that spin transport and spin Hall effect occur in different parts of the same material gives rise to a hitherto unreported efficiency for the spin-to-charge voltage output. Remarkably for a single graphene/MoS$_2$ heterostructure-based device, we evidence a superimposed spin-to-charge current conversion that can be indistinguishably associated with either the proximity-induced Rashba-Edelstein effect in graphene or the spin Hall effect in MoS$_2$. By comparing our results to theoretical calculations, the latter scenario is found the most plausible one. Our findings pave the way towards the combination of spin information transport and spin-to-charge conversion in two-dimensional materials, opening exciting opportunities in a variety of future spintronic applications.**

The efficient transport and the manipulation of spins in the same material are mutually exclusive as they would require simultaneously weak and strong spin-orbit coupling (SOC) respectively. Graphene, due to its low intrinsic SOC, is proven to be an outstanding material that can transport spins over long distance of tens of micrometres[1-10]. For the same reason, the generation and tuning of spin currents in graphene are out of reach, limiting its capability to active spintronic device functionalities and related applications. To solve this issue, methods to artificially induce SOC in graphene have been explored. For instance, the SOC in graphene has been enhanced by chemical doping[11-17] or by proximity-induced coupling with materials possessing large SOC[18-35]. The latter method is more convenient since the chemical properties of graphene are not altered, whereas its high-quality electronic transport properties are preserved.

Transition metal dichalcogenides (TMDs) with chemical formula MX$_2$ (M=Mo, W and X=S, Se) are layered materials of semiconducting nature displaying unique combined electronic, optical, spintronic and valleytronic properties[36-50]. They possess a strong intrinsic SOC of tens of meV, few orders larger than that of pristine graphene[36,37]. Accordingly, a large intrinsic spin Hall effect (SHE) has been theoretically predicted in TMDs[38]. However, its experimental observation remains elusive because of the technical difficulties to inject and detect spin information into these materials[49,50]. Only recently, a spin-orbit torque experiment has shown a large spin-to-charge conversion in monolayer $MoS_2$ and $WSe_2$ interfaced with a ferromagnet, although it has been attributed to Rashba-Edelstein effect (REE)[51]. Interestingly, TMDs can be hybridized to graphene forming van der Waals heterostructres[18-35] Theoretical calculations show that TMDs imprint their strong SOC into graphene via proximity effect, with an induced SOC of the order of few meV[17,18]. The strong induced SOC in TMD-proximitized graphene has been confirmed by many weak anti-localization measurements[21-24]. Another predicted feature of this proximity effect is the induced spin-valley coupling in graphene, leading to a large spin lifetime anisotropy[25] which has also been recently confirmed experimentally[26,27].

Further interesting theoretical predictions for TMD-proximitized graphene include the emergence of the SHE[29,30] and the Rashba-Edelstein effect (REE)[30,31]. The SHE and the REE are exotic phenomena that convert charge currents into spin currents (in case of SHE)[52-56] or spin densities (in case of REE)[57-62]. Their reciprocal effects [inverse SHE (ISHE) and inverse REE (IREE)] convert spin currents or spin densities into charge currents. Since these mechanisms lead to highly efficient electrical generation and detection of spins, they can play an important role towards the realization of all-electrically controlled spintronic devices, such as spin-orbit torque memories[63-65] or spin-orbit logic circuits[66,67]. Whereas the SHE has been mostly studied in heavy metals with strong SOC[52-56], the REE has been observed in two-dimensional electron gas systems[58-60] as well as interfaces lacking inversion symmetry[61,62]. In graphene, previous measurements of the SHE using nonlocal Hall bar geometries have been reported[15,16,20]. However, a variety of effects unrelated to spin can contribute to such nonlocal measurements, questioning the interpretation of the results and therefore impeding a further optimization and implementation of those spin effects into practical applications[68-72]. A convincing experimental proof of spin-to-charge conversion due to the SHE or the REE in graphene is thus crucially lacking. In this letter, by carefully designing graphene/TMD-based lateral spin valves (LSVs), we solve this long-standing experimental challenge, and report an unprecedented manifestation of spin-to-charge conversion in graphene/$MoS_2$ van der Waals heterostructures that can be unambiguously attributed to the SHE in graphene. Additionally, a second contribution, which can be induced by either the REE in graphene or the SHE in $MoS_2$ is also clearly observed. Our theoretical calculations indicate that the latter is the most plausible source, thus providing the first all-electrical measurement of spin Hall effect in $MoS_2$. Realization of spin transport and spin-to-charge conversion in the same material (in this case graphene) a long awaiting milestone for the field of spintronics.

Our device design consists of $MoS_2$ placed in the middle of cross-shaped graphene as shown in Figure 1. This design enables studying spin-charge interconversion stemming from three distinct sources: 1) Proximity-induced SHE in graphene[30,53] (Figure 1a); 2) Proximity-induced REE in graphene[30,31,62] (Figure 1b); 3) SHE in $MoS_2$ [55,56,76,77] (Figure 1c). For the first case, a charge current applied along the graphene/$MoS_2$ stripe results in the generation of spin current with out-of-plane spin polarization along the graphene channel. In the second and third cases, a charge current applied along the graphene/$MoS_2$ stripe gives rise to a spin current with the same in-plane polarization along the graphene channel. The signal induced by the SHE in graphene can thus be separated from the other two effects by verifying the spin polarization directions. The Onsager reciprocal equivalents, ISHE and IREE, can also be studied and separated in the same manner.

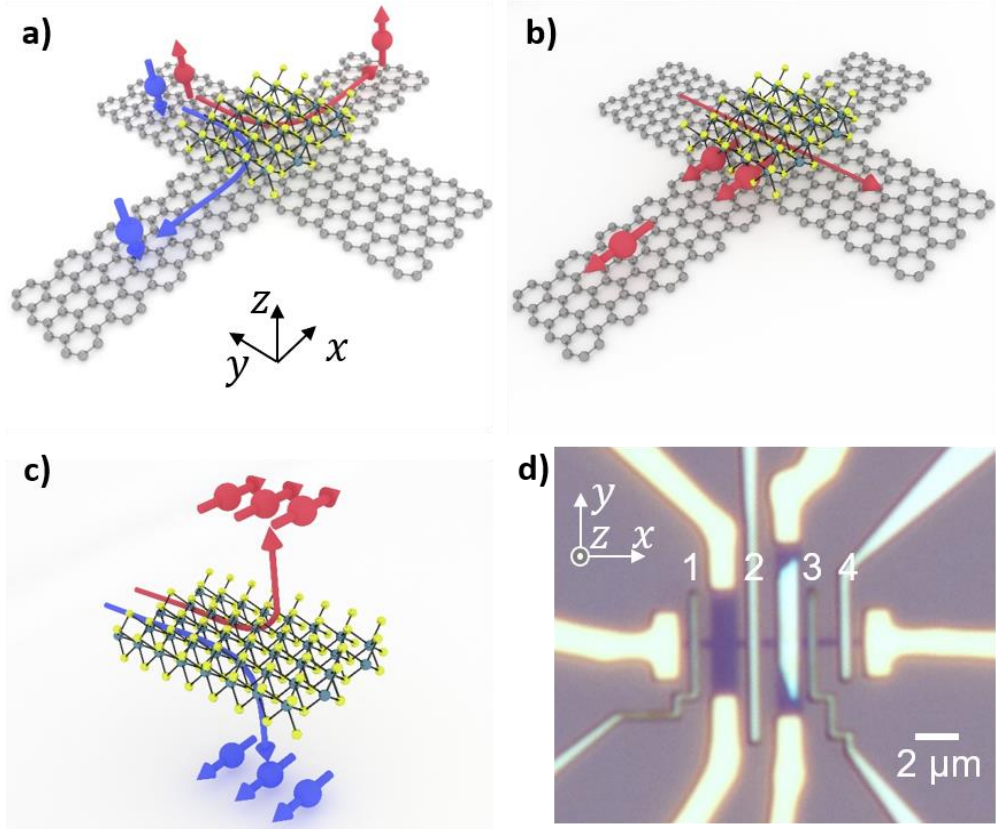

**Figure 1.** Description of our device configuration and possible spin-charge interconversion scenarios. **a)** Sketch of proximity-induced spin Hall effect in graphene. A charge current applied along the graphene/MoS$_2$ stripe ($y$-axis) results in a spin current with out-of-plane (along $z$) spin polarization in the graphene channel along $x$. **b)** Sketch of proximity-induced Rashba-Edelstein effect in graphene. A charge current applied along the graphene/MoS$_2$ stripe ($y$-axis) creates a spin accumulation with in-plane (along $x$) spin polarization which then diffuses to the graphene channel along $x$. **c)** Sketch of spin Hall effect in MoS$_2$. A charge current applied along the $y$-axis of MoS$_2$ creates an out-of-plane (along $z$) spin current with in-plane (along $x$) spin polarization. When this MoS$_2$ is placed on top of graphene, this spin current diffuses to the graphene channel along $x$. **d)** Optical microscope image of one of our devices (sample A). It contains graphene shaped into two Hall bars connected each other (blue). The ends of the graphene stripes are connected to Ti/Au contacts (yellow). The MoS$_2$ flake (light blue) lies on one of the graphene Hall bars. Four Co/TiO$_x$ electrodes (grey) are placed on top of the graphene channel.

To study and quantify the different spin-to-charge conversion contributions, we fabricated devices as the one shown in Figure 1d (sample A). Each device contains exfoliated few-layer graphene shaped into a narrow channel ($W_{gr}$ = 350 nm) with a double Hall bar ($W_H$ = 1.2 μm). A multilayer MoS$_2$ flake lies on top of one of the graphene Hall bars. Four ferromagnetic (FM) Co electrodes (with interface resistances ranging from 10 to 70 kΩ due to a TiO$_x$ underlayer) are placed on the graphene channel forming three different LSVs. A detailed description of the device fabrication process is given in methods. As shown in Figure 1d, the LSV with FM electrodes 1 and 2 and the LSV with electrodes 3 and 4 can both be used as reference LSVs. The LSV with FM electrodes 2 and 3 contains MoS$_2$ placed on top of the graphene Hall bar between the FM electrodes. The spin-to-charge conversion is measured between the graphene/MoS$_2$ vertical stripe and FM electrode 3.

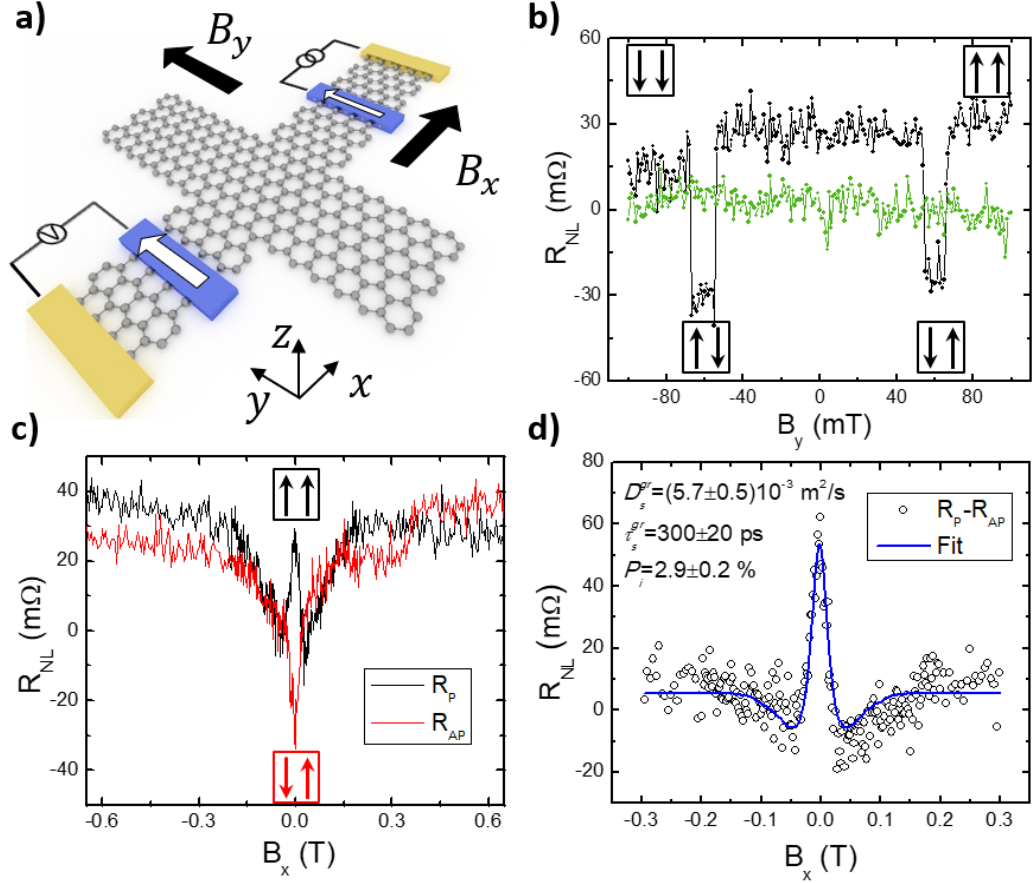

**Figure 2.** Spin transport in LSVs. **a)** Sketch of the nonlocal resistance measurement at the reference graphene LSV. **b)** Nonlocal resistance measurements performed by applying field along the in-plane easy axis ($B_y$). The black and green curves correspond to the reference graphene LSV (with FM electrodes 1 and 2) and the LSV with MoS$_2$ in between (with FM electrodes 2 and 3), respectively. Baselines are subtracted in both curves. **c)** Nonlocal symmetric Hanle curves measured at the reference graphene LSV by applying a magnetic field along the in-plane hard axis ($B_x$) for the initial parallel ($R_P$, black) and antiparallel ($R_{AP}$, red) configuration of the FM electrodes. Baselines are subtracted in both curves. **d)** Spin signal calculated using $R_P - R_{AP}$ from the Hanle data in c) (black open circles) with the corresponding fit (blue solid line) using the Bloch equations and the extracted parameters. Data in all panels correspond to sample A at 10 K. An electrical current $I_C = 10$ μA is used.

First, we study the reference LSV with FM electrodes 1 and 2 to extract the spin transport properties of pristine graphene. The measurements are done in the standard nonlocal configuration[73,74]. A charge current ($I_C$) is injected from a Co electrode into the graphene channel, creating a spin accumulation at the Co/graphene interface. This spin accumulation diffuses toward both sides of the graphene channel, creating a pure spin current out of the current path, which is detected by another Co electrode as a nonlocal voltage ($V_{NL}$), see Figure 2a. The nonlocal resistance $R_{NL} = V_{NL}/I_C$ is high ($R_{NL}^P$, parallel) or low ($R_{NL}^{AP}$, antiparallel) depending on the relative orientation of the magnetization of the two electrodes, which can be set by applying an in-plane magnetic field in the $y$–direction ($B_y$) due to the different shape anisotropies of the electrodes (Figure 2b). The spin signal, which is defined as $\Delta R_{NL} = R_{NL}^P - R_{NL}^{AP}$, is ~60 mΩ at 10 K. Moreover, a Hanle precession measurement has been performed to characterize the spin transport properties of the graphene channel (Figure 2c). Since the injected spins are oriented along the $y$–direction, a perpendicular in-plane magnetic field $B_x$ is applied. The precession and decoherence of the spins cause the oscillation and decay of the signal. In addition, the rotation of the Co magnetizations with $B_x$ tends to align the polarization of the injected spin current with the applied field, restoring the $R_{NL}$ signal to its zero-field value when the Co electrodes reach parallel magnetizations along the $x$–direction at high enough $B_x$. By the proper combination of Hanle measurements for initial parallel and antiparallel states of the Co electrodes, the contribution from spin precession and decoherence (Figure 2d) and the rotation angle $\beta$ of the Co magnetization (Note S1 in the Supporting Information) can be obtained. The Hanle curve for the spin precession and decoherence is then fitted to the solution of the Bloch equations[75] as shown in Figure 2d (for details, see Note S1 in the Supporting Information). From this fitting, the spin lifetime of the pristine graphene $\tau_s^{gr}$=300±20 ps, the spin polarization of the Co/graphene interface $P_i$=2.9±0.2% and the spin diffusion constant of the pristine graphene $D_s^{gr}$=(5.7±0.5)× $10^{-3}$ m²/s are extracted. Next, we show the nonlocal resistance in the LSV with the middle MoS$_2$ flake (between FM electrodes 2 and 3). $\Delta R_{NL}$ in this case is expected to be lower than that of the reference LSV ($\Delta R_{NL}^{ref}$) due to the well-known spin absorption effect[33,34] and the proximity-induced spin relaxation in the graphene channel[26,27,32]. Depending on the quality of the graphene/MoS$_2$ interface, the reduction of the spin signal can vary. In our case, the measured spin signal for graphene/MoS$_2$ LSV at 10 K is smaller than the noise level (~5 mΩ), as shown in Figure 2b.

Once we confirmed that the spin signal diminishes at the graphene/MoS$_2$ region, we inspect the possibility of spin-to-charge conversion in the same region. For this, we initially align the magnetization of the Co electrodes along the easy axis by applying a magnetic field ($\pm B_y$). Subsequently, while sweeping the field along the in-plane hard axis ($\pm B_x$), $V_{NL}$ is measured across the graphene/MoS$_2$ bar (Figure 3d and 3f) while injecting a constant current $I_C$=10 μA from the Co electrode 3 to the graphene channel. As mentioned above, a voltage across the graphene/MoS$_2$ branch placed along the $y$–direction can be obtained either due to the spin-to-charge conversion of i) out-of-plane spins with spin polarization along $z$ or ii) in-plane spins with spin polarization along $x$. For the first case, as the initial spin polarization ($y$) and the applied field ($x$) are orthogonal, the spins precess in the $y - z$ plane around the field. This process results in an antisymmetric Hanle signal exhibiting either a maximum or minimum at certain values of $\pm B_x$, when the spins reaching the graphene/MoS$_2$ region point out of plane ($z$). The antisymmetric Hanle curve is also reversed when the initial magnetization direction is switched, since the injected spins are opposite (Figure 3d)[30,46,77]. For the second case, as the required spin polarization and the applied field are along the same direction ($x$), $R_{NL}$ increases or decreases while the electrode magnetization rotates towards $\pm x$ and saturates at values of $\pm B_x$ that correspond to the saturation magnetization of the Co electrode (Figure 3f). This results in an S-shaped signal that is odd with $B_x$ and independent of the initial magnetization direction of the Co electrode[76,77]. As shown in Figure 3a, we clearly observe antisymmetric Hanle curves which reverse by switching the initial magnetization direction. This confirms that, at the graphene/MoS$_2$ region, the out-of-plane spins are converted into a charge current, which in open circuit condition gives rise to $V_{NL}$. As explained in Figure 1, the only possible source of out-of-plane spin-to-charge conversion in our experimental design is the ISHE in graphene due to the SOC induced by MoS$_2$ (Figure 1a). Therefore, we unambiguously conclude the observation of proximity-induced ISHE in graphene in sample A. For further confirmation, we repeat the same experiment in a different sample (sample B) obtaining similar results. In sample B, we also performed measurements where the spin current direction is reversed by injecting current with the Co electrode on the right and the left side of the graphene/MoS$_2$ region, respectively (Note S3 in the Supporting Information). We find that the spin-to-charge conversion signal changes sign with reversing the spin current direction, which further confirms the proximity-induced ISHE in graphene as the source of the signal, since the ones generated by the other possible sources (REE in graphene and SHE in MoS$_2$) do not depend on the spin current direction in the graphene channel.

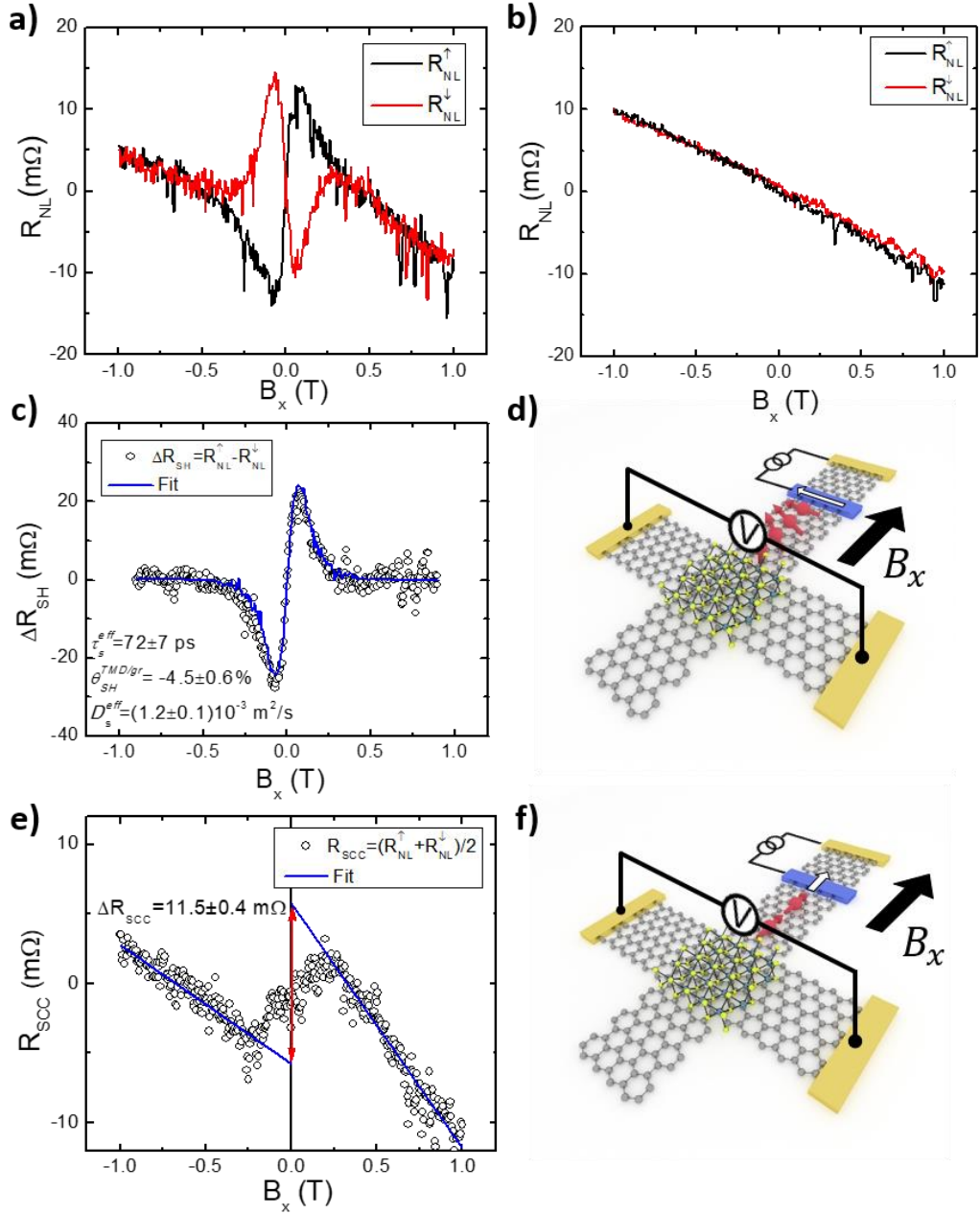

**Figure 3.** Spin-to-charge conversion measurement and analysis. **a)** Nonlocal spin-to-charge conversion curves obtained by applying a charge current between Co electrode 3 and the right Ti/Au electrode and measuring the voltage across the graphene/MoS$_2$ stripe. The magnetic field is applied along the in-plane hard axis direction ($B_x$) for initial positive (black) and negative (red) magnetization directions of the Co electrodes. **b)** Same measurements obtained by applying a charge current between Co electrode 2 and the right Ti/Au electrode and measuring the voltage across the pristine graphene stripe in the reference LSV. **c)** Net antisymmetric Hanle signal (open circles) obtained by subtracting the two Hanle curves ($\Delta R_{SH} = R_{NL}^\uparrow - R_{NL}^\downarrow$) in a), which is fitted with the solution of the Bloch equation (blue curve) and the extracted parameters. **d)** Sketch of the measurement configuration that results in the antisymmetric Hanle curve shown in c). When $B_x$ is low enough, the Co magnetization is in its easy axis and the injected (red) spins are polarized along $y$, leading to a spin precession along the $y$–$z$ plane. **e)** Background signal (open circles) extracted by averaging the two Hanle curves [$R_{SCC} = (R_{NL}^\uparrow + R_{NL}^\downarrow)/2$] in a). The magnitude of this spin-to-charge conversion signal ($\Delta R_{SCC}$) is quantified by calculating the zero-field extrapolation using linear fittings to the signal at high positive and negative fields. **f)** Sketch of the measurement configuration that results in the S-shaped curve shown in e). When $B_x$ saturates the Co magnetization in its hard axis, the injected (red) spins are polarized along $x$ and do not precess. Data in all panels correspond to sample A at 10 K.

To quantify the overall spin-to-charge conversion due to the ISHE in proximitized graphene, the two antisymmetric Hanle curves for the initial positive and the negative magnetizations direction of the Co electrodes were subtracted ($\Delta R_{SH} = R_{NL}^{\uparrow} - R_{NL}^{\downarrow}$) and the resulting curve is shown in Figure 3c. Fitting this resulting curve to the solution of the Bloch equation (Note S2 in the Supporting Information), we extract a spin Hall angle $\theta_{SH}^{TMD/gr}= -4.5\pm0.6\%$, an effective spin lifetime $\tau_s^{eff}=72\pm7$ ps, and an effective spin diffusion constant $D_s^{eff}=(1.2\pm0.1)\times 10^{-3}$ m$^2$/s at 10 K. Comparing the variation of $V_{NL}$ with respect to the Co magnetization direction and the direction of $B_x$, the spin Hall angle is found to be negative. The robustness of this value is confirmed in sample B, where $\theta_{SH}^{TMD/gr}= -4.8\pm0.9\%$, $\tau_s^{eff}=70\pm9$ ps, and $D_s^{eff}=(1.5\pm0.1)\times 10^{-3}$ m$^2$/s are obtained at 10 K.

These values are an order of magnitude larger than $\theta_{SH}^{TMD/gr}$~0.1% predicted theoretically for MoS$_2$ (Ref. 29). We understand such difference by noting that those calculations have been limited to a phenomenological description of disorder using intrinsic broadening of the electronic states. Here by performing spin Hall conductivity calculations for monolayer graphene/MoS$_2$ heterostructures, we traced back the origin of the SHE to the combination of the valley-Zeeman interaction, the intrinsic SOC, and the staggered potential. Using this model, and taking into account the broadening effects and temperature, we predict a $\theta_{SH}^{TMD/gr}$ ranging from 1 to 10%, depending on the position of the Fermi Level. $\theta_{SH}^{TMD/gr}$ is larger close to the charge neutrality point and changes from negative to positive when going from the valence to the conduction band. This last result would then imply that, in the current experiment, graphene is p-doped. Details are explained in Note S8.1 in the Supporting Information.

To confirm that the spin-to-charge conversion occurs strictly in the graphene regions in proximity with MoS$_2$, and not in the pristine graphene itself, we perform a control experiment in the reference LSV by injecting current with Co electrode 2 and measuring $V_{NL}$ across the vertical pristine graphene stripe placed next to it (Figure 1d). Figure 3b shows no evidence of Hanle spin precession effect. Instead, the fact that $R_{NL}$ varies linearly with the magnetic field confirms that the spin-to-charge conversion measurement in graphene/MoS$_2$ is not due to any experimental spurious effect. Ideally, we expect the linear background in the reference LSV to be also present in the background of the antisymmetric Hanle curve (Figure 3a) measured in graphene/MoS$_2$. To check this, we extract the background signal of the spin-to-charge conversion measurements in graphene/MoS$_2$ by excluding the antisymmetric component [$R_{SCC} = (R_{NL}^{\uparrow} + R_{NL}^{\downarrow})/2$] as shown in Figure 3e (see Note S4.1 in the Supporting Information). Surprisingly, we found an S-shaped background signal of amplitude 11.5 m$\Omega$ with changing slopes at magnetic field $B_x \sim \pm240$ mT, the same field that saturates the magnetization of the Co electrodes. This indicates that an in-plane spin-to-charge conversion signal in graphene/MoS$_2$ is also measured.

As explained above, there are two possible sources for in-plane spin-to-charge conversion in our device configuration: the proximity-induced IREE in graphene (Figure 1b) and the ISHE in MoS$_2$ (Figure 1c). In our experiment, both are indistinguishable, jeopardizing a separate quantification of the efficiencies of the two effects. However, we can separately calculate the limits of the efficiencies of each phenomenon by assuming that the signal is entirely obtained from a single effect. For the proximity-induced IREE in graphene, we quantify the spin-to-charge conversion efficiency to be $\alpha_{RE}\sim$ +0.85% (a detailed derivation is given in Note S4.2 in the Supporting Information). One of the major differences between the REE and the SHE is that the former produces a non-equilibrium spin density instead of a spin current (as is the case of the latter), which complicates the definition of a figure of merit. An equivalent of a spin Hall angle $\alpha_{RE}$ can be nevertheless proposed by assuming that the spin accumulation transforms entirely into a spin current by diffusion, and divide it by the injected charge current density. To obtain such information, numerical simulations based on Kubo methods are performed[29,30] together with the analytical theory of Offidani et al.[31]. Using the same spin-orbit parameters used to compute $\theta_{SH}^{TMD/gr}$, we estimate that the REE dimensionless efficiency $\alpha_{RE}$ would be at maximum 0.1% (see Note S8.2 in the Supporting Information for details), which is one order of magnitude smaller than the value extracted from the experiment. More remarkably, the theoretical analysis predicts that the sign of the proximity-induced REE is the same as for the proximity-induced SHE. Since $\theta_{SH}^{TMD/gr}$ is negative, $\alpha_{RE}$ should also be negative. Therefore, the positive sign of the in-plane spin-to-charge conversion rules out the proximity-induced REE as its origin.

On the other hand, if the in-plane spin-to-charge conversion occurs due to ISHE in $MoS_2$, it requires spin absorption from graphene into $MoS_2$, which depends on the resistivities and spin diffusion lengths of both materials and their interface resistance [33,34]. As graphene and $MoS_2$ are stamped together in our device, it is not possible to separately quantify the resistivity of $MoS_2$ ($\rho_{MoS_2}$). Therefore, we examine the possible strength of ISHE in $MoS_2$ by varying $\rho_{MoS_2}$ (see Note S4.3 in the Supporting Information). Here, the spin-to-charge conversion voltage measured across the graphene/$MoS_2$ channel depends on two competing contributions that vary in opposite way with $\rho_{MoS_2}$: 1) an increase in $\rho_{MoS_2}$ decreases the spin absorption and, thus, the spin-to-charge conversion voltage; 2) the effective resistance of the graphene/$MoS_2$ region increases with $\rho_{MoS_2}$ and, thus, increases the spin-to-charge conversion voltage until it saturates due to the shunting effect of the graphene channel. In the optimal resistivity value ($\rho_{MoS_2} \sim 7 \times 10^{-4}$ Ωm), a minimum spin Hall angle of $MoS_2$ $\theta_{SH}^{MoS_2} \sim 3.3\%$ is required to obtain the spin-to-charge conversion signal of 11.5 mΩ. In this scenario, we can estimate a spin Hall conductivity $\sigma_{SH}^{MoS_2} \sim 0.47$ $\Omega^{-1}cm^{-1}$, which is reasonable for a bulk n-doped $MoS_2$ flake, as theoretically calculated in Ref. 38. It thus seems more plausible that the observed in-plane spin-to-charge conversion arises from the ISHE in $MoS_2$.

We finally performed additional spin-to-charge conversion measurements at 100 K, 200 K and 300 K (see Notes S5 and S6 in the Supporting Information). As shown in Figure 4a for sample A, the out-of-plane component of the spin-to-charge conversion signal clearly decreases with increasing temperature, while the in-plane component persists up to room temperature, indicating that the former is strongly influenced by the temperature in contrast to the latter. Unfortunately, we could not measure for both magnetization directions of the Co injector, which prevents us from a full quantification of the ISHE in the proximitized graphene as done at 10 K. For sample B, we could properly extract the spin Hall angle, which strongly decays with temperature from –4.8±0.9% at 10 K down to –0.33±0.04% at 300 K (Figure 4b). For this estimation, we also need to take into account the temperature dependence of the graphene/$MoS_2$ sheet resistance (inset in Figure 4b), which we observe to decrease with increasing temperature, a feature reported in graphene when the Fermi level is near the Dirac point[77]. Theoretically, the spin Hall conductivity of proximitized graphene is also expected to decrease with temperature close to the neutrality point due to the change of sign when crossing from the valence to the conduction band. In Note S8.1 in the supporting information, we show that the spin Hall conductivity is inversely proportional to the temperature, which combined with the decrease of the sheet resistance with temperature leads to a fast decay of the spin Hall angle.

Concerning the in-plane spin-to-charge conversion, it can be quantified up to room temperature in sample A because in this case one curve at a single magnetization direction of the Co injector already gives reliable information. This component slightly decreases from 11.5 mΩ at 10 K to 7.4 mΩ at 300 K. On the one hand, if we assume the origin of the signal is IREE in proximitized graphene, the conversion efficiency $\alpha_{RE}$ increases from 0.85±0.03% at 10 K to 3.0±0.2% at 300 K.

This is in sharp contrast with our theoretical calculations, which predict that $\alpha_{RE}$ strongly decays with temperature due to the antisymmetric behavior between the valence and conduction bands which will be mixed in the presence of temperature (see Note S8.2 in the Supporting Information for details). On the other hand, if we assume the origin of the signal is ISHE in MoS$_2$, we need a minimum $\theta_{SH}^{MoS_2}$ ~2.2% to explain the signal at 300 K, a value close to the ~3.3% at 10 K. Such a weak temperature variation in the spin Hall angle is expected for a bulk material, which is the case of our MoS$_2$ flake. Although we cannot fully rule out any scenario, the temperature dependence of the in-plane spin-to-charge conversion suggests again that spin absorption and subsequent ISHE in the MoS$_2$ flake is a more plausible option. It is worth mentioning that no clear in-plane spin-to-charge conversion is observed in sample B (see Notes S4.1 and S5 in the Supporting Information). The lack of signal in sample B can be again better understood if one assumes that its origin is the ISHE in MoS$_2$, rather than the IREE in proximitized graphene: whereas the spin absorption into MoS$_2$ depends on details of the graphene and the MoS$_2$ flake that are independent of the presence of the ISHE in the proximitized graphene, IREE in proximitized graphene should be present whenever the proximity-induced ISHE occurs.

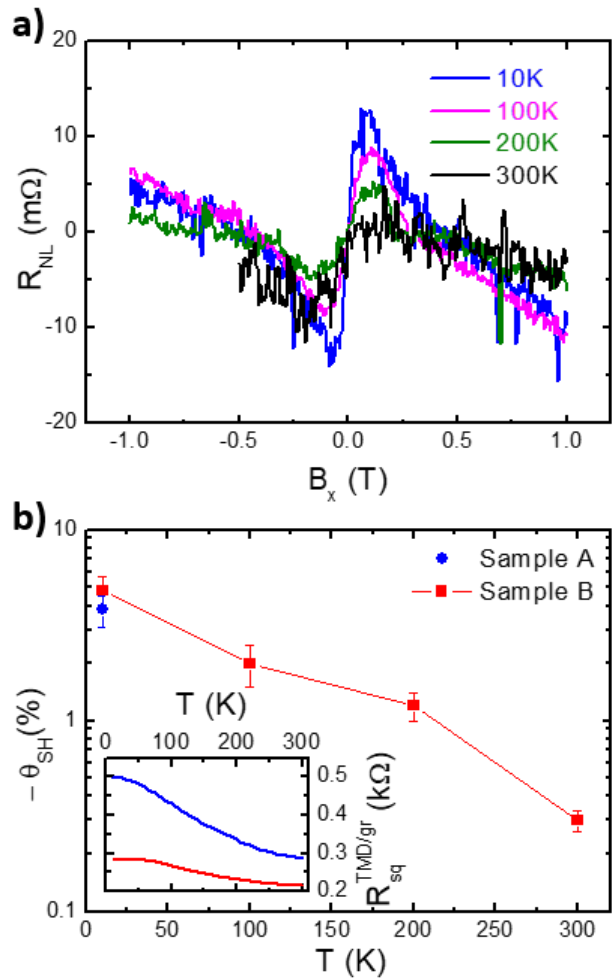

**Figure 4.** Temperature dependence of the spin-to-charge conversion. **a)** Nonlocal spin-to-charge conversion curves for initial positive magnetization direction of Co electrode 3 at different temperatures in sample A. **b)** Spin Hall angle of the proximitized graphene at different temperatures for sample A (blue solid circle) and sample B (red solid squares). **Inset:** Square resistance as a function of temperature from two-point measurements along the graphene/MoS$_2$ stripe in samples A and B.

Finally, it is also worth noting that, although the spin Hall angle in proximitized graphene is smaller than that of the best SHE materials such as heavy metals or topological insulators, the overall spin-to-charge conversion efficiency can be very high. A proper way to quantify this efficiency is the ratio $R_{eff}$ between the output voltage ($\Delta V_{SH}$) and the injected spin current reaching the spin Hall region ($I_S$), which has units of resistance (see Note S7 in the Supporting Information). In the device geometry used in this work, $R_{eff}^{TMD/gr} = 2\theta_{SH}^{TMD/gr} R_{sq}^{TMD/gr} (\bar{I}_s/I_s)$, where $\bar{I}_s$ is the average spin current at the spin Hall region. Since all parameters are known, we can estimate $R_{eff}^{TMD/gr}$ = 13.4 Ω. Previous reports of the largest spin-to-charge conversion output were performed in graphene/Pt LSVs[76,77] in which graphene was used to transport spins into Pt where the spin-to-charge conversion occurs. In this case, the efficiency is defined as $R_{eff}^{Pt} = (2\theta_{SH}^{Pt} \rho_{Pt} x_{Pt/gr}/W_{Pt})(\bar{I}_s/I_s)$, where $\theta_{SH}^{Pt}$, $\rho_{Pt}$, $W_{Pt}$ are the spin Hall angle, resistivity and width of Pt, and $x_{Pt/gr}$ a correction factor that considers the current in Pt shunted through the graphene channel. Taking the parameters from Ref. 76, we obtain $R_{eff}^{Pt/gr}$=0.27 Ω. In other words, the unique feature of spin transport and spin-to-charge conversion in graphene itself (with $\theta_{SH}^{TMD/gr}$ = − 4.5%) results in an efficiency 50 times larger than when spin-to-charge conversion occurs in a different material after spin absorption, even if Pt (with $\theta_{SH}^{Pt} = 23\%$) is used.

To conclude, an unprecedented and unambiguous experimental demonstration of the proximity-induced ISHE has been found in graphene together with a manifestation of another spin-to-charge conversion phenomenon either due a proximity-induced IREE in graphene or an ISHE in MoS$_2$, being the latter more likely. We obtained the largest spin-to-charge conversion efficiency as compared to previous reports in LSVs. The device concept proposed here, a LSV with a cross-shaped graphene channel, can easily integrate a gate voltage which is expected to tune the spin-to-charge conversion in graphene[29,30], enabling an extra functionality with a lot of potential which is not possible in prototypical spin Hall metals. The device concept can also be extended to study spin-to-charge conversion in a variety of material combinations of graphene with heavy metals, oxides, different TMDs, and topological insulators, which have been also suggested to be suitable companions for generating giant REE[79]. Most importantly, our proof of concept in which spins can be generated, transported and detected in the same material (graphene) is a propitious feature for developing future spintronics devices.

**Methods.** *Device fabrication.* The graphene/MoS$_2$ van der Waals heterostructures are fabricated by mechanical exfoliation followed by dry viscoelastic stamping. We first exfoliate graphene from bulk graphitic crystals (supplied by NGS Naturgraphit GmbH) using a Nitto tape (Nitto SPV 224P) onto Si substrates with 300 nm SiO$_2$. Few-layer graphene flakes are identified by optical contrast under an optical microscope. Then a MoS$_2$ crystal (supplied by SPI supplies) is exfoliated using the Nitto tape and transferred on to a piece of poly-dimethyl siloxane (Gelpak PF GEL film WF 4, 17 mil.). After identifying a short and narrow MoS$_2$ flake using an optical microscope, it is stamped on top of graphene using visco-elastic stamping tool where a three-axes micrometer stage is used to accurately position the two flakes. The flake is then structured into graphene double cross-shape using electron-beam lithography followed by reactive ion etching. After etching, the sample is annealed for 1 h at 400 °C in ultrahigh vacuum ($10^{-9}$ torr) to possibly remove the contamination. The graphene is then connected with Ti(3nm)/Au(40nm) contacts fabricated using electron-beam lithography followed by electron beam deposition in ultrahigh vacuum and lift-off. Using the same procedure, the 35 nm thick Co electrodes are fabricated on top of the graphene channel. Before this deposition, a TiO$_x$ tunnel barrier is fabricated by depositing 3Å of Ti and subsequent natural oxidation in air. The widths of the Co electrodes vary between 250 nm to 400 nm leading to different coercive fields for each electrode. The exact dimensions of the devices are extracted from scanning electron and atomic force microscopy after they have been measured (Note S9 in the Supporting Information).

*Electrical measurements.* Charge and spin transport measurements are performed in a Physical Property Measurement System by Quantum Design, using a 'DC reversal' technique with a Keithley 2182 nanovoltmeter and a 6221 current source at temperatures ranging from 10 to 300 K. We apply in-plane and out-of-plane magnetic fields with a superconducting solenoid magnet and a rotatable sample stage.

**Acknowledgements**

This work is supported by the Spanish MINECO under the Maria de Maeztu Units of Excellence Programme (MDM-2016-0618) and under Projects MAT2015-65159-R and MAT2017-82071-ERC, and by the European Union H2020 under the Marie Curie Actions (QUESTECH). S.R and J.H.G acknowledge support from the European Union Seventh Framework Programme under Grant Agreement No. 785219 Graphene Flagship and the computational resources from PRACE and the Barcelona Supercomputing Center (Project No. 2015133194).

# SUPPORTING INFORMATION

## S1. Analysis of the conventional (symmetric) Hanle precession measurements

### S1.1. Standard model

To extract the spin transport parameters of the pristine graphene channel and determine the spin injection efficiency of the ferromagnetic (FM) electrodes, we have performed conventional symmetric Hanle precession experiments in the pristine graphene region. These are performed by measuring the nonlocal resistance of the reference graphene lateral spin valve (LSV) while having a parallel ($R_P$) or antiparallel ($R_{AP}$) orientation of the FM electrode magnetizations and applying a magnetic field in the $x$ direction ($B_x$) (i.e., perpendicular to the easy axis of the FM), as shown in Figure S1a for the case of Sample A at 10 K. Finally, the spin transport parameters have been obtained from fitting to the following equation [S1]:

$$\Delta R_{NL} = R_P - R_{AP} = \frac{P_{sym}^2 \cos^2(\beta) R_{sq}^{gr} \lambda_s^{gr}}{W_{gr}} Re\left\{\frac{e^{-\frac{L}{\lambda_s^{gr}}\sqrt{1-i(\omega-\omega_0)\tau_s^{gr}}}}{\sqrt{1-i(\omega-\omega_0)\tau_s^{gr}}}\right\} + R_0 \qquad (S1)$$

Here, $P_{sym}^2 = P_i P_d$ is the product of the injector and detector spin polarization respectively, $\beta$ is the angle between the contact magnetization and the easy axis [see Figure S1b], $R_{sq}^{gr}$ is the square resistance of the graphene channel, $\lambda_s^{gr} = \sqrt{D_s^{gr} \tau_s^{gr}}$ is the spin relaxation length, $D_s^{gr}$ is the spin diffusion coefficient, $\tau_s^{gr}$ is the spin lifetime, $W_{gr}$ is the channel width, $\omega = g\mu_B B_x/\hbar$ is the Larmor frequency, $g = 2$ is the Landé factor, $\mu_B$ is the Bohr magneton, and $\hbar$ is the reduced Plank constant. Finally, $\omega_0 = g\mu_B B_0/\hbar$ accounts for a small remanescense from the magnet and $R_0$ is the background signal. Equation S1 assumes that spin transport is 1D and the contact resistances are much higher than the channel spin resistance ($R_N^{gr} = R_{sq}^{gr} \lambda_s^{gr}/W_{gr}$) [S2].

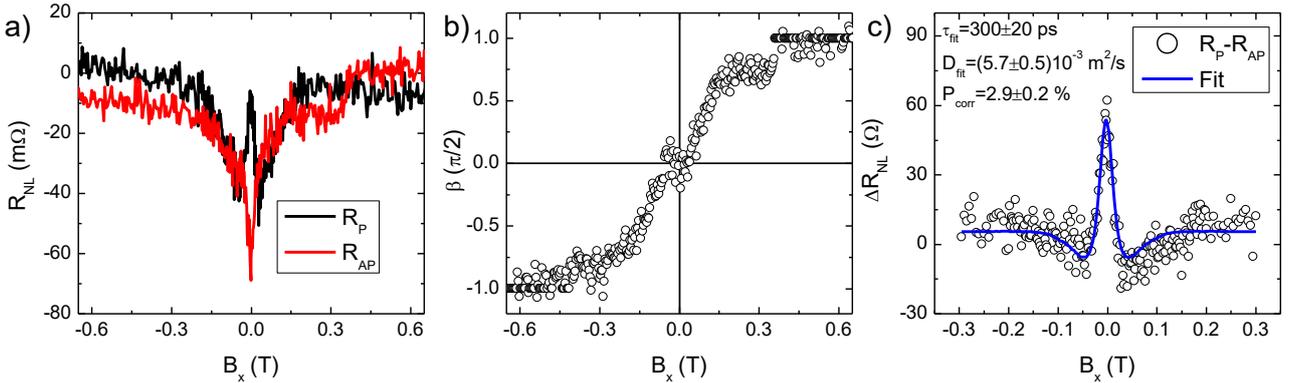

**Figure S1.** (a) Conventional Hanle precession data in the parallel and antiparallel contact magnetization configurations for sample A at 10 K. (b) Angle between the contact magnetization and the easy axis extracted from the Hanle data in a) as described in the text below. (c) Spin signal calculated using $R_P - R_{AP}$ from the Hanle data in (a) with the corresponding fit using Eq. S1 and extracted parameters. $P_{corr}$ is the spin polarization obtained using the model with an extra arm described in Section S2 and the fit parameters as inputs.

Since $B_x$ is applied perpendicular to the easy axis of a FM, it pulls its magnetization by an angle $\beta$ in the field direction [S3]. To determine $\beta$, we use the fact that the measured nonlocal resistance [Figure S1a] does not include only the precessing component described above, but also includes a signal which is generated by the spins which are injected parallel to the $B_x$ direction as the contact magnetizations are pulled. This term is proportional to $\sin^2(\beta)$ and appears in both $R_P$ and $R_{AP}$ with the same sign. Because the precessing component has opposite sign for $R_P$ and $R_{AP}$, $R_P + R_{AP}$ is proportional to $\sin^2(\beta)$. To obtain the data plotted in Figure S1b, we have normalized $0 < R_P + R_{AP} < 1$, and taken the arcsine of its square root to obtain $\beta$ as a function of the magnetic field. Finally, to guarantee that the extracted $\beta$ at $B_x = 0$ is zero and that it reaches $\pi/2$ at high $B$, we have renormalized the result of this operation. At $B = 0.4$ T, we see a jump in $\beta$, which we attribute to a switch of one of the magnetizations. To prevent this step from affecting our analysis, we have used only the negative $B$ result for the fits, such as the one shown in Figure S1c where the spin signal $R_P - R_{AP}$ is fit to Eq. S1. The obtained fit parameters are shown in the blue rows of Table S1 (sample A) and Table S2 (sample B).

In the case of sample A, the reference graphene LSV has two extra arms (the Hall bar) connected to the main channel, which implies that spins are also diffusing towards these arms. To account for the effect of the arms on the spin polarization we have assumed that spin propagation is 1D and the arms are much longer than the spin relaxation length. As a consequence, the shape of the Hanle curve (determined by $D_s^{Gr}$ and $\tau_s^{Gr}$) is not affected within our model by the presence of the extra arms and only the spin injection and detection efficiencies have to be adjusted. The spin transport parameters extracted from the fit are used as inputs to the three-arm model described in Section S1.2 to determine a better estimate of the contact spin polarization, which we call $P_{corr}$.

For Sample B, the symmetric Hanle curves were taken in a reference LSV without any cross between spin injector and detector, hence, $P_i$ is extracted directly from the fit of $\Delta R_{NL} = R_P - R_{AP}$ to Eq. S1.

**S1.2. Three-arm model**
The original geometry we want to model for the reference graphene LSV used in Sample A is shown in Figure S2a.

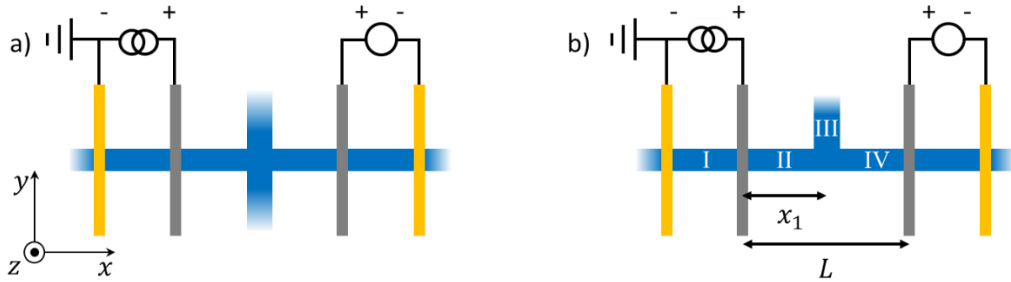

**Figure S2**. (a) 1D approximation of the reference device geometry for Sample A. (b) Simulated device geometry. The width of the top arm is two times the width of the arms in the device and in (a) and the white numbers correspond to the different modelling regions (see discussion below).

However, because we are interested in the spins that diffuse towards the middle arm and the spin transport is fully diffusive, the two arms can be simplified down to one with twice the width [see Figure S2b]. The most important assumptions here are 1D spin transport and non-invasive contacts. The left contact is the injector and the right one the detector.

To calculate the spin signal, one needs to define the spin accumulation, which is a solution of the Bloch equations considering a magnetic field applied in the $z$ direction [S4]:

$$D_s^{gr} \frac{d^2 \mu_x}{dx^2} - \frac{\mu_x}{\tau_s^{gr}} - \omega \mu_y = 0$$

$$D_s^{gr} \frac{d^2\mu_y}{dx^2} - \frac{\mu_y}{\tau_s^{gr}} + \omega\mu_x = 0$$

Note that here we have considered that the magnetic field is applied in the $z$ direction for analogy with the conventional Hanle precession, but we are applying a magnetic field along $x$. We have done this because the solution for $\mu_y$ is the same in both cases for an isotropic system such as pristine graphene. The general solution for $\mu_y$ for each region can be written as:

$$\mu_I^y = Ae^{\frac{x}{\lambda_s^{gr}}\sqrt{1+i\omega\tau_s^{gr}}} + Be^{\frac{x}{\lambda_s^{gr}}\sqrt{1-i\omega\tau_s^{gr}}}$$

$$\mu_{II}^y = Ce^{\frac{x}{\lambda_s^{gr}}\sqrt{1+i\omega\tau_s^{gr}}} + De^{\frac{x}{\lambda_s^{gr}}\sqrt{1-i\omega\tau_s^{gr}}} + Ee^{-\frac{x}{\lambda_s^{gr}}\sqrt{1+i\omega\tau_s^{gr}}} + Fe^{-\frac{x}{\lambda_s^{gr}}\sqrt{1-i\omega\tau_s^{gr}}}$$

$$\mu_{III}^y = Ge^{-\frac{x}{\lambda_s^{gr}}\sqrt{1+i\omega\tau_s^{gr}}} + He^{-\frac{x}{\lambda_s^{gr}}\sqrt{1-i\omega\tau_s^{gr}}}$$

$$\mu_{IV}^y = Ie^{-\frac{x}{\lambda_s^{gr}}\sqrt{1+i\omega\tau_s^{gr}}} + Je^{-\frac{x}{\lambda_s^{gr}}\sqrt{1-i\omega\tau_s^{gr}}}$$

From the Bloch equations we obtain an expression for $\mu_x$:

$$\mu_x = -\frac{D_s^{gr}}{\omega}\frac{d^2\mu_y}{dx^2} + \frac{\mu_y}{\omega\tau_s^{gr}}$$

which results, at the different regions,

$$\mu_I^x = -iAe^{\frac{x}{\lambda_s^{gr}}\sqrt{1+i\omega\tau_s^{gr}}} + iBe^{\frac{x}{\lambda_s^{gr}}\sqrt{1-i\omega\tau_s^{gr}}}$$

$$\mu_{II}^x = -iCe^{\frac{x}{\lambda_s^{gr}}\sqrt{1+i\omega\tau_s^{gr}}} + iDe^{\frac{x}{\lambda_s^{gr}}\sqrt{1-i\omega\tau_s^{gr}}} - iEe^{-\frac{x}{\lambda_s^{gr}}\sqrt{1+i\omega\tau_s^{gr}}} + iFe^{-\frac{x}{\lambda_s^{gr}}\sqrt{1-i\omega\tau_s^{gr}}}$$

$$\mu_{III}^x = -iGe^{-\frac{x}{\lambda_s^{gr}}\sqrt{1+i\omega\tau_s^{gr}}} + iHe^{-\frac{x}{\lambda_s^{gr}}\sqrt{1-i\omega\tau_s^{gr}}}$$

$$\mu_{IV}^x = -iIe^{-\frac{x}{\lambda_s^{gr}}\sqrt{1+i\omega\tau_s^{gr}}} + iJe^{-\frac{x}{\lambda_s^{gr}}\sqrt{1-i\omega\tau_s^{gr}}}$$

To determine the spin accumulation in the system we need to determine the coefficients $A - H$. This is done by applying the following boundary conditions. First, we impose the continuity of $\mu^x$ and $\mu^y$ at $x = 0$ (intersection point between regions I and II) and $x = x_1$ (intersection point between regions II, III, and IV). From these conditions, we obtain 6 equations. Second, we define the spin current as $I_s^i = -\frac{W_{gr}}{eR_{sq}^{gr}}\frac{d\mu_s^i}{dx}$ where $i$ refers to the spin polarization direction of the spin accumulation ($x$ or $y$). Next, we impose the continuity of $I_s^x$ and $I_s^y$ at $x = x_1$ and the continuity of $I_s^x$ and discontinuity of $I_s^y$ of $P_iI_c$, where $P_i$ is the injector polarization and $I_c$ the applied current, at $x = 0$. This gives the 4 equations that we are missing.

Finally, the solution for the spin signal is:

$$\Delta R_{NL} = R_P - R_{AP} = \frac{2P_d\mu_{IV}^y(L)}{eI_c} = \frac{2P_d}{eI_c}\left(Ie^{-\frac{L}{\lambda_s^{gr}}\sqrt{1+i\omega\tau_s^{gr}}} + Je^{-\frac{L}{\lambda_s^{gr}}\sqrt{1-i\omega\tau_s^{gr}}}\right)$$

In the final solution, both $I$ and $J$ are proportional to $P_i$ and, hence, the spin signal is proportional to $P_i P_d = P_{corr}^2$.

Here, we use the above equation together with the spin transport parameters in the pristine graphene region obtained from the fit to Eq. S1 and the magnitude of the measured spin signal to obtain $P_{corr}$, that is the average spin polarization of the ferromagnetic contacts.

## S2. Analysis of the (antisymmetric) Hanle precession measurements with spin Hall detection

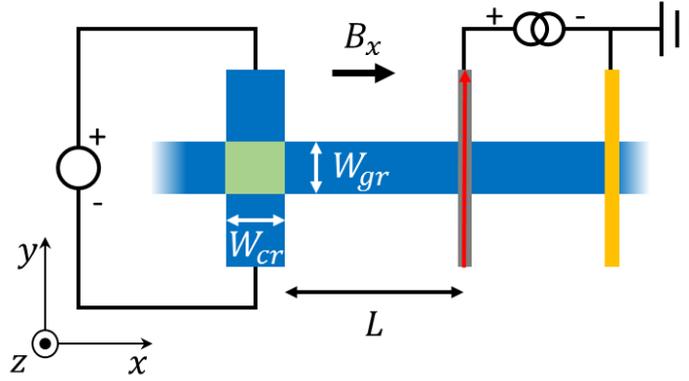

**Figure S3**. Sketch of the measured geometry with the positive magnetic field and the "up" direction of the magnetization indicated with arrows. The green area is covered by a TMD and is where the spin-to-charge conversion takes place.

When a spin current density $I_s/W_{gr}$ enters a region with high spin-orbit coupling, it generates a transverse charge current density $I_c^T/W_{cr} = \theta_{SH}^{TMD/gr} I_s/W_{gr}$ via the inverse spin Hall effect, here $\theta_{SH}$ is the spin Hall angle and $W_{cr}$ the length of the high spin-orbit coupling region (see Figure S3). This generates a transverse voltage $V_{SH}$

$$\frac{V_{SH}}{W_{cr}R_{gr}} = \frac{I_c^T}{W_{cr}} = \frac{\theta_{SH}^{TMD/gr} I_s}{W_{gr}} \rightarrow V_{SH} = \frac{\theta_{SH}^{TMD/gr} R_{gr} W_{cr} I_s}{W_{gr}}$$

where $R_{gr} = R_{sq}^{TMD/gr} W_{gr}/W_{cr}$ is the resistance for the transverse current that determines the conversion from an induced current to a voltage. Note that $R_{sq}^{TMD/gr}$ is the square resistance of the TMD-covered graphene region. Therefore,

$$V_{SH} = \frac{\theta_{SH}^{Gr} R_{sq}^{TMD/gr} W_{gr} W_{cr} I_s}{W_{cr} W_{gr}} = \theta_{SH}^{Gr} R_{sq}^{TMD/gr} I_s$$

To obtain the spin currents propagating in our device, we assume that it is homogeneous and isotropic, and the propagation is 1D. Because the transport properties of the TMD-covered region are not the same as those of the pristine graphene region, we will extract effective parameters ($\tau_s^{eff}$, $D_s^{eff}$, and $\lambda_s^{eff}$) from the homogeneous fit. These parameters are an average for both the pristine and the TMD-covered graphene regions.

The out-of-plane ($z$ direction) spin accumulation ($\mu_z$) at a distance $L$ from the spin injector, which induced by spin precession around a magnetic field applied in the $x$ direction is written as [S4, S5]:

$$\mu_z = \pm \frac{P_i e I_c R_{sq}^{gr} \lambda_s^{eff}}{2W_{gr}} Im\left\{\frac{e^{-\frac{L}{\lambda_s^{eff}}\sqrt{1-i\omega\tau_s^{eff}}}}{\sqrt{1-i\omega\tau_s^{eff}}}\right\}$$

Here, $P_i$ the spin injection efficiency of the injector, $I_c$ the bias current. The $\pm$ corresponds to the up and down direction of the injector magnetization defined as in Figure S3.

The spin current is defined as:

$$I_s = I_\uparrow - I_\downarrow = \frac{W_{gr}}{eR_{sq}^{Gr}}\frac{d\mu_z}{dx} = \pm\frac{P_i I_c}{2} Im\left\{e^{-\frac{L}{\lambda_s^{eff}}\sqrt{1-i\omega\tau_s^{eff}}}\right\}$$

Then the signal induced by this spin current is:

$$R_{SH} = \frac{V_{SH}}{I_c} = \frac{\theta_{SH}^{Gr} R_{sq}^{TMD/gr} I_s}{I_c} = \pm\frac{\theta_{SH}^{Gr} P_i R_{sq}^{TMD/gr}}{2} Im\left\{e^{-\frac{L}{\lambda_s^{eff}}\sqrt{1-i\omega\tau_s^{eff}}}\right\}$$

However, in our geometry, we have a cross with a width $W_{cr}$ (Figure S3) that is comparable to the spin relaxation length. Therefore, $I_s$ needs to be replaced by the average spin current $\bar{I}_s = \frac{1}{W_{cr}}\int_L^{L+W_{cr}} I_s dx$. The result from this operation is:

$$R_{SH} = \frac{\theta_{SH}^{TMD/gr} R_{sq}^{TMD/gr} \bar{I}_s}{I_c} = \pm\frac{\theta_{SH}^{TMD/gr} P_i R_{sq}^{TMD/gr} \lambda_s^{eff}}{2W_{cr}} Im\left\{\frac{e^{-\frac{L}{\lambda_s^{eff}}\sqrt{1-i\omega\tau_s^{eff}}}}{\sqrt{1-i\omega\tau_s^{eff}}} - \frac{e^{-\frac{L+W_{cr}}{\lambda_s^{eff}}\sqrt{1-i\omega\tau_s^{eff}}}}{\sqrt{1-i\omega\tau_s^{eff}}}\right\}$$

To remove any signal which does not come from the $y$-component of the injector magnetization in our measurement we subtract both injector configurations to

$$\Delta R_{SH} = R_{SH}^\uparrow - R_{SH}^\downarrow = \frac{P_{anti}^2 R_{sq}^{TMD/gr} \lambda_s^{eff}}{W_{cr}} Im\left\{\frac{e^{-\frac{L}{\lambda_s^{eff}}\sqrt{1-i(\omega-\omega_0)\tau_s^{eff}}}}{\sqrt{1-i(\omega-\omega_0)\tau_s^{eff}}} - \frac{e^{-\frac{L+W_{cr}}{\lambda_s^{eff}}\sqrt{1-i(\omega-\omega_0)\tau_s^{eff}}}}{\sqrt{1-i(\omega-\omega_0)\tau_s^{eff}}}\right\} + R_0 \quad (S2)$$

where $R_{SH}^\uparrow$ and $R_{SH}^\downarrow$ correspond to the spin Hall component of the nonlocal resistance of the graphene/TMD lateral device (Figure S3) while orienting the FM injector along the $+y$ and $-y$ direction, respectively, and applying $B$ in the $x$ direction, which leads to antisymmetric Hanle precession curves. The measured signal with these injector configurations is called $R_{nl}^\uparrow$ and $R_{nl}^\downarrow$. Note that we have also added the term $\cos(\beta)$ to account for the effect of the magnetic field on the magnetization direction of the FM injector. Equation S2 has been used to fit to the antisymmetric Hanle precession curves such as the ones shown in Figure S4. From the fit we obtain the effective

$\tau_s^{eff}$, $D_s^{eff}$ and $P_{anti}$ of the channel, which are averaged over the pristine graphene and TMD-covered regions. Note that $P_{anti} = \sqrt{P_i \theta_{SH}^{TMD/gr}}$ depends on both the injector spin polarization $P_i$ and the spin Hall angle $\theta_{SH}^{TMD/gr}$. To determine $\theta_{SH}^{TMD/gr}$ we assume that $P_i = P_{corr}$ for Sample A and $P_i = P_{sym}$ for sample B obtained from the symmetric Hanle curve (Section S1). The fitted parameters are shown in the orange rows of Table S1 (Sample A) and Table S2 (Sample B).

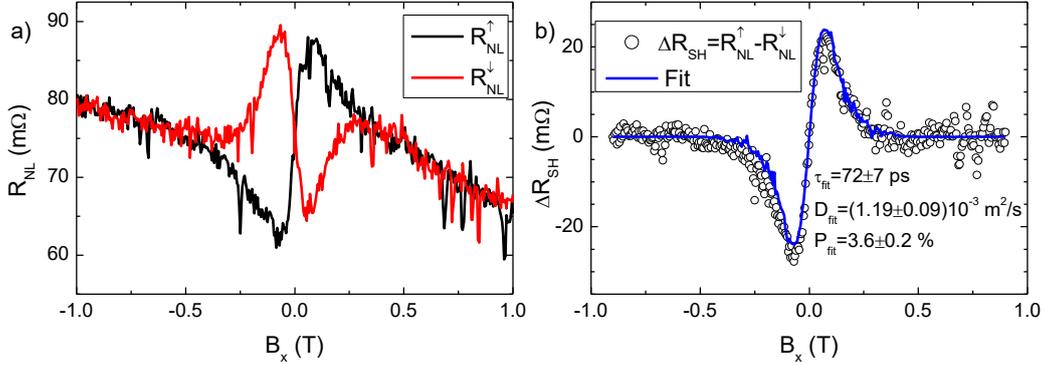

**Figure S4.** (a) Antisymmetric Hanle precession measurements for the two different orientations of the FM injector for Sample A at 10 K. Note that we have called $R_{NL}^\uparrow$ and $R_{NL}^\downarrow$ the signals measured with the magnetization of the spin injector aligned in the $+y$ and $-y$ direction respectively. (b) $\Delta R_{SH}$ and its fit to Eq. S2 with the extracted parameters.

## S3. Measurement of spin Hall signals obtained with the spin injector placed at the right and left side of the Hall cross

Since the spin signal is induced by the spin current, which points in the graphene plane, it should change sign when the spin injector is changed from the right to the left side of the Hall cross. Note that, if the spin signal would be induced by spin absorption in the TMD, it would not change sign. This control experiment has been performed in sample B, where the two measurements shown in Figure S5 confirm that the signal is induced by the spin current propagating in the graphene plane.

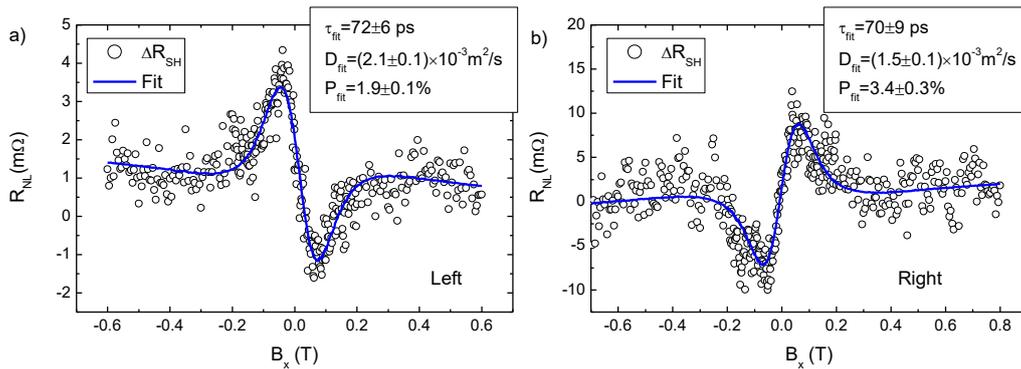

**Figure S5.** Antisymmetric Hanle curves obtained by subtracting the measurements obtained for both magnetic configurations of the FM injector at 10 K in sample B, for the (a) left and (b) right spin injectors.

We see that, as expected, the spin transport parameters ($\tau_{fit}$ and $D_{fit}$) extracted from both Hanle curves are very similar. However, because the magnitude of the spin signals is different, the spin polarizations extracted from both curves show a discrepancy of almost a factor of two. This is caused by the fact that, in both cases we are using different spin injectors which have different efficiencies. In Table S2, this has consequences on the determination of $\theta_{SH}^{TMD/gr}$. To avoid this discrepancy, we rely on the measurements performed using the right injector, in which the spin injector is the same as for the reference Hanle curves.

## S4. Analysis of the $\mu_x$-induced spin-to-charge conversion signal

### S4.1. Extraction of the signal

As explained above, when a magnetic field $B_x$ is applied perpendicular to the easy axis of a FM electrode, its magnetization rotates an angle $\beta$ in the direction of the field. In our geometry, this results in the injection of spins in the direction of the magnetic field independently of the initial magnetization direction and, therefore, this component can be extracted calculating $R_{SCC} = (R_{NL}^\uparrow + R_{NL}^\downarrow)/2$. The result from this operation is shown in Figure S6 for the case of Sample A at 10 K.

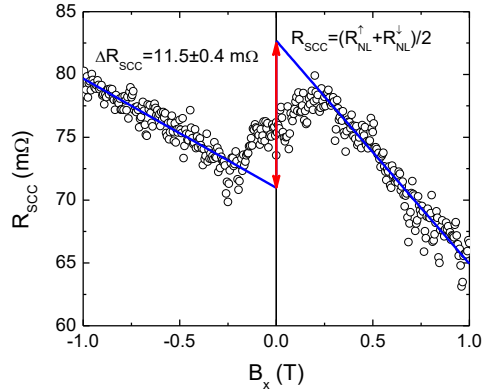

**Figure S6.** $\mu_x$-induced spin-to-charge conversion signal $R_{SCC}$ obtained by averaging the antisymmetric Hanle precession curves measured for both magnetic configurations of the FM injector at 10 K for Sample A [Figure S4a]. The magnitude of the spin-to-charge conversion signal ($\Delta R_{SCC}$) is quantified by calculating the zero-field extrapolation using linear fittings to the spin signal at high positive and negative fields.

The S-shaped signal saturates at around ±240 mT, which is the saturation field for the magnetization of the FM injector [see Figure S1b]. This result indicates that the source of this signal depends on the $x$ component of the spin accumulation ($\mu_x$). Because there is a background which has different slope for positive and negative magnetic fields, we extrapolate the high $B$ data to $B_x = 0$ to extract the magnitude of this component. As discussed in the main manuscript, this means that it can be caused either by the Rashba-Edelstein effect in the graphene layer or the spin Hall effect in the MoS$_2$ semiconducting channel after spin absorption.

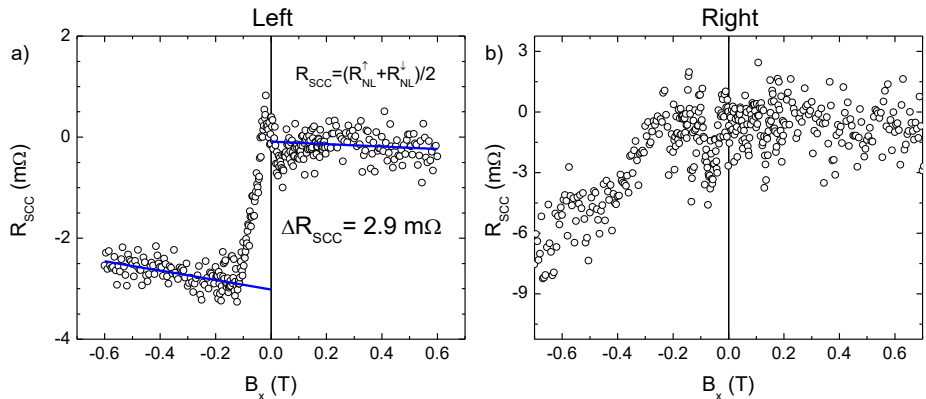

**Figure S7.** $R_{SCC}$ signal at 10 K for Sample B measured with spins injected by a FM electrode placed on (a) the right and (b) the left side of the TMD-covered region.

We have performed the same operation on the measurements obtained from Sample B and the results at 10 K are plotted in Figure S7. In contrast with the left injector, the right one does not show any $\mu_x$-induced spin-to-charge conversion feature.

### S4.2. Quantification of the Rashba-Edelstein effect

For the Rashba-Edelstein case, a spin accumulation in the x direction ($\mu_x$) induces a charge current density in the y direction $J_c^T = I_c^T/W_{cr}$ with an efficiency $\gamma$ that is the conversion efficiency: $\mu_x \gamma = \frac{I_c^T}{W_{cr}} = \frac{V_{RE}}{W_{cr} R_{gr}}$. As a consequence, $V_{RE} = \overline{\mu_x} \gamma R_{sq} W_{gr}$ where $\bar\mu_x = \frac{1}{W_{cr}} \int_L^{L+W_{cr}} \mu_x(x) dx$ is the average spin accumulation at the cross, which is defined as $\mu_x = \pm \frac{e P_i I R_{sq}^{TMD/gr} \lambda_s^{eff}}{2 W_{gr}} e^{-\frac{x}{\lambda_s^{eff}}}$ where + and − correspond to the positive and negative magnetization directions of the spin injector corresponding to positive and negative magnetic field respectively.

$$\bar\mu_x = \frac{1}{W_{cr}} \int_L^{L+W_{cr}} \mu_x(x) dx = \pm \frac{e P_i I_c R_{sq}^{TMD/gr} \lambda_s^{eff}}{2 e W_{gr}} \frac{1}{W_{cr}} \int_L^{L+W_{cr}} e^{-x/\lambda_s^{eff}} dx$$

$$= \pm \frac{e P_i I_c R_{sq}^{TMD/gr} \lambda_s^{eff}}{2 W_{gr}} \frac{1}{W_{cr}} \left[-\lambda_s^{eff} e^{-\frac{x}{\lambda_s^{eff}}}\right]_L^{L+W_{cr}}$$

$$= \pm \frac{e P_i I_c R_{sq}^{TMD/gr} \lambda_s^{eff\,2}}{2 W_{gr} W_{cr}} \left(e^{-\frac{L}{\lambda_s^{eff}}} - e^{-\frac{L+W_{cr}}{\lambda_s^{eff}}}\right)$$

$$R_{RE} = \frac{V_{RE}}{I_c} = \frac{\overline{\mu_x} \gamma R_{sq} W_{gr}}{I_c} = \pm \frac{\gamma R_{sq} W_{gr}}{I_c} \frac{e P_i I R_{sq} \lambda_s^{eff\,2}}{2 W_{gr} W_{cr}} \left(e^{-\frac{L}{\lambda_s^{eff}}} - e^{-\frac{L+W_{cr}}{\lambda_s^{eff}}}\right)$$

$$= \pm \frac{e P_i \gamma R_{sq}^2 \lambda_s^{eff\,2}}{2 W_{cr}} \left(e^{-\frac{L}{\lambda_s^{eff}}} - e^{-\frac{L+W_{cr}}{\lambda_s^{eff}}}\right)$$

$$\Delta R_{RE} = R_{RE}^{\rightarrow} - R_{RE}^{\leftarrow} = \frac{e P_i \gamma R_{sq}^2 \lambda_s^{eff\,2}}{W_{cr}} \left(e^{-\frac{L}{\lambda_s^{eff}}} - e^{-\frac{L+W_{cr}}{\lambda_s^{eff}}}\right) \tag{S3}$$

We also want to quantify the Rashba-Edelstein effect using a dimensionless parameter. For this purpose, we assume that the spin currents are the source of the transverse charge currents and not the spin accumulations. For this, we use the formula derived in [S6].

$$\Delta R_{RE} = R_{RE}^{\rightarrow} - R_{RE}^{\leftarrow} = \frac{\alpha_{RE} P_i R_{sq} \lambda_s^{eff}}{W_{cr}} \left(e^{-\frac{L}{\lambda_s^{eff}}} - e^{-\frac{L+W_{cr}}{\lambda_s^{eff}}}\right) \tag{S4}$$

Finally, to obtain an estimation for $\gamma$ and $\alpha_{RE}$, we assume that the entire observed signal arises from Rashba-Edelstein effect, i.e., $\Delta R_{SCC} = \Delta R_{RE}$. The extracted parameters from these operations are shown in the green rows of Table S1 for Sample A. For Sample B we did only see a clear signal for one of the configurations at 10 K.

### S4.3. Quantification of the spin Hall effect in MoS$_2$

Because of the reduced spin lifetime of the MoS$_2$-covered graphene region it is not straightforward to determine the spin current which is absorbed by the semiconductor. However, we can estimate an upper limit of the signal expected if the spin current diffusing from the graphene channel into the MoS$_2$ is converted to a charge current due to the spin Hall effect in the MoS$_2$ channel.

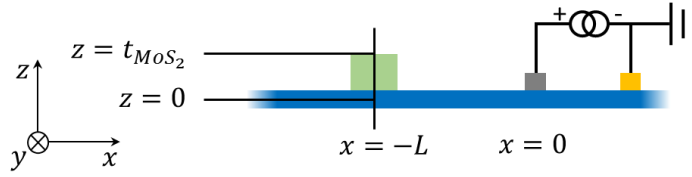

**Figure S8**. Sketch of the modeled geometry. The spin injector is grey, the Ti/Au contact is yellow, the graphene channel is blue, and the MoS$_2$ is green. The thickness of the graphene is assumed to be zero.

To evaluate the effect of spin absorption in a 1D approach which we can solve analytically, we assume that the width of the MoS$_2$ flake is significantly smaller than the spin relaxation length in the graphene channel ($W_{cr} \ll \lambda_s^{gr}$). We also ignore the role of the cross and assume that the graphene channel is 1D through all its length. With these assumptions, the modeled geometry is shown in Figure S8.

To model for this geometry, we break the device into different regions: I. The right side of the spin injector. II. From $x = 0$ to $x = -L$, III. MoS$_2$, from $z = 0$ to $z = t_{MoS_2}$, and IV. Graphene for $x < -L$. We write the spin accumulation in the different parts of the sample as follows:

$$\mu_I = A e^{-\frac{x}{\lambda_s^{gr}}}$$
$$\mu_{II} = B e^{\frac{x}{\lambda_s^{gr}}} + C e^{-\frac{x}{\lambda_s^{gr}}}$$
$$\mu_{III} = D e^{\frac{z}{\lambda_s^{MoS_2}}} + E e^{-\frac{z}{\lambda_s^{MoS_2}}}$$
$$\mu_{IV} = F e^{\frac{x}{\lambda_s^{gr}}}$$

where $\lambda_s^{MoS_2}$ is the spin relaxation length at the MoS$_2$ and we have imposed that $\mu_s(x \to \pm\infty) \to 0$. To obtain the spin accumulation in the system we need to determine the coefficients $A - F$. For this purpose, we need 6 equations. We obtain three of them from imposing the continuity of $\mu_{IV}$ at $x = 0$ and $x = -L$.

To obtain the other 3 equations, we need to define the spin currents in the graphene and MoS$_2$ channels respectively:

$$I_s^{gr} = -\frac{W_{gr}}{eR_{sq}^{gr}} \frac{d\mu_s^{gr}}{dx}$$
$$I_s^{MoS_2} = -\frac{W_{gr} W_{cr}}{e\rho_{MoS_2}} \frac{d\mu_s^{MoS_2}}{dz}$$

Here, $\rho_{MoS_2}$ is the resistivity of the MoS$_2$ channel, and $W_{cr}$ is the width of the MoS$_2$ flake in Figure S8. By imposing the discontinuity of $I_s^{gr}$ at $x = 0$, which is of $P_i I_c$ as discussed in section S2, the continuity of $I_s^{gr}$ and $I_s^{MoS_2}$ at $x = L$, and $I_s^{MoS_2}(z = t_{MoS_2}) = 0$, that guarantees that there is no spin current leaving the MoS$_2$ flake, we obtain the remaining 3 equations.

From the derivation above, we obtain the spin currents and accumulations in the MoS$_2$ channel. However, we still need to convert them into voltages to determine the measured signal. A spin current density in the MoS$_2$ channel ($J_s^{MoS_2} = \frac{I_s^{MoS_2}}{W_{gr}W_{cr}}$) induces a transverse charge current density $J_c^T = \frac{I_c^T}{W_{cr}t_{MoS_2}} = \theta_{SH}^{MoS_2} \frac{I_s^{MoS_2}}{W_{gr}W_{cr}}$ where $\theta_{SH}^{MoS_2}$ is the spin Hall angle of the MoS$_2$.

Because we are measuring the voltage in an open circuit configuration, we use: $V_{SH}^{MoS_2} = I_c^T R_{eff}$, where $R_{eff}$ is the effective resistance of the MoS$_2$ and graphene. Because the resistance of the MoS$_2$ is considerably higher than that of the graphene channel, we need to take the effect of shunting by the graphene into account to determine $R_{eff}$. We assume that the system behaves like if the MoS$_2$ channel would be connected in parallel with the graphene:

$$R_{eff}^{-1} = R_{MoS_2}^{-1} + R_{gr}^{-1} = \frac{W_{cr}t_{MoS_2}}{W_{gr}\rho_{MoS_2}} + \frac{W_{cr}}{R_{sq}W_{gr}}$$

Finally,

$$V_{SH}^{MoS_2} = I_c^T R_{eff} = \theta_{SH}^{MoS_2} \frac{I_s^{MoS_2} t_{MoS_2} R_{eff}}{W_{gr}}$$

Because the spin current is not constant along the MoS$_2$ thickness we replace $I_s^{MoS_2}$ in the equation above by:

$$\bar{I}_s^{MoS_2} = \frac{1}{t_{MoS_2}} \int_0^{t_{MoS_2}} I_s^{MoS_2}(z)dz = \frac{W_{cr}W_{gr}}{et_{MoS_2}\rho_{MoS_2}}\left[D\left(1 - e^{t_{MoS_2}/\lambda_s^{MoS_2}}\right) + E\left(1 - e^{-t_{MoS_2}/\lambda_s^{MoS_2}}\right)\right]$$

Using the two equations above, we determine the voltage induced by the spin Hall effect in MoS$_2$, which we divide by the current applied to the spin injector to obtain the nonlocal resistance:

$$R_{SH}^{MoS_2} = \frac{V_{SH}^{MoS_2}}{I_C} = \theta_{SH}^{MoS_2} \frac{\bar{I}_s^{MoS_2} t_{MoS_2} R_{eff}}{I_c W_{gr}}$$

As explained above, the sign of $R_{SH}^{MoS_2}$ changes with the magnetization direction. Because our definition of the spin-to-charge conversion signal is $\Delta R_{SCC} = R_{SCC}^{\rightarrow} - R_{SCC}^{\leftarrow}$, we define $\Delta R_{SH}^{MoS_2} = 2R_{SH}^{MoS_2}$. The inputs of our model are $\lambda_s^{MoS_2}$, $\rho_{MoS_2}$, and $\theta_{SH}^{MoS_2}$ and are unknown to us, while the output is $\Delta R_{SH}^{MoS_2}$ and we know its value. To give an estimate for $\theta_{SH}^{MoS_2}$, we assume that $\lambda_s^{MoS_2} = 20$ nm, as estimated in Ref. [S7], we leave $\rho_{MoS_2}$ as a free parameter and extract the $\theta_{SH}^{MoS_2}$ that gives a signal $\Delta R_{SH}^{MoS_2} = \Delta R_{SCC}$. The results from this model are shown in Figure S9 for sample A at 10 K. Because of the effect of shunting and the reduction of the spin currents which get absorbed by the MoS$_2$ as $\rho_{MoS_2}$ increases, $R_{SH}^{MoS_2}$ presents a maximum value of 11.5 mΩ at $\rho_{MoS_2} = 7.1 \times 10^{-4}$ Ωm. This helps us to give an estimate of the best-case scenario for spin absorption that requires the lowest $\theta_{SH}^{MoS_2}$ to achieve the measured signal. We have adjusted $\theta_{SH}^{MoS_2}$ so that the maximum $R_{SH}^{MoS_2}$ corresponds to the measured signal. From this process we conclude that the minimal spin Hall angle required to achieve $\Delta R_{SCC}$ at 10 K in sample A is 3.3%. To estimate the temperature dependence of $\theta_{SH}^{MoS_2}$ for sample A, we have performed the same analysis described above assuming that the spin relaxation length in MoS$_2$ does not depend on the temperature, which is the case if the Dyakonov-Perel mechanism is the dominant source of spin relaxation [S8]. The results are shown in Table S1.

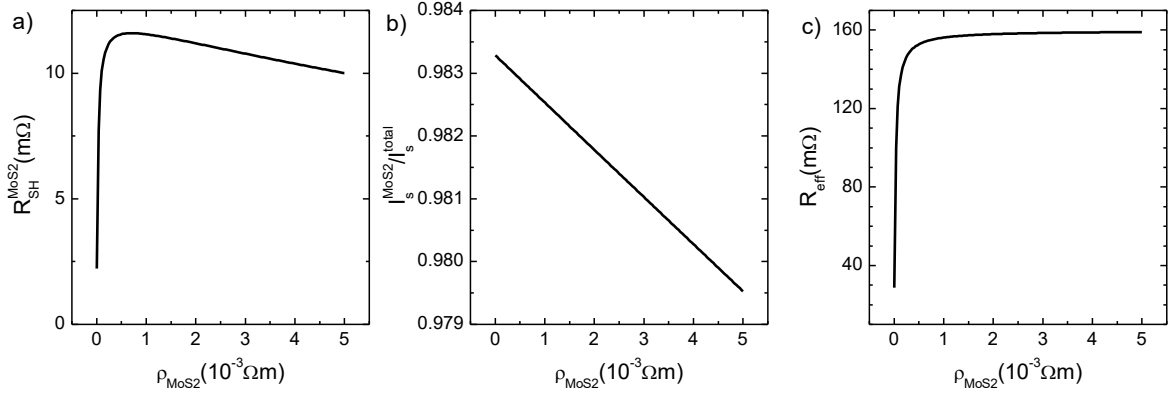

**Figure S9**. (a) Estimation of the spin signal induced by the spin Hall effect in the MoS$_2$ channel as a function of the MoS$_2$ resistivity using the parameters of Sample A at 10 K and by assuming $\theta_{SH}^{MoS_2} = 3.3$ %. (b) Ratio between the spin current diffusing into the MoS$_2$ ($I_s^{MoS_2}$) and spin current entering the graphene-MoS$_2$ junction in the model ($I_s^{total}$). The ratio decreases as $\rho_{MoS_2}$ increases. (c) Effective resistance of the MoS$_2$ and graphene channel. As $\rho_{MoS_2}$ increases, $R_{eff}$ increases before saturating when the MoS$_2$ resistance becomes significantly higher than that of the graphene channel.

## S5. Analysis of the measurements at different temperatures

### S5.1. Sample A
Unlike the case of 10 K (Fig. 3a of the main text), we only have one magnetic injector configuration for the antisymmetric Hanle precession curves ($R_{NL}^\uparrow$) measured at 100 K, 200 K, and 300 K (Figure 4a of the main text). Hence, we cannot quantify the spin Hall angle in graphene for Sample A at these temperatures. In contrast, the $\mu_x$-induced spin-to-charge conversion signals can be obtained from direct extrapolation of $R_{NL}^\uparrow$ from high fields, as explained in section S3.1. The extracted signal $\Delta R_{SCC}$ is shown in Table S1.

### S5.2. Sample B
For sample B, we have both magnetic configurations of the antisymmetric Hanle curves ($R_{NL}^\uparrow$ and $R_{NL}^\downarrow$) at 100, 200 and 300 K and, consequently, $\Delta R_{SH} = R_{NL}^\uparrow - R_{NL}^\downarrow$ can be fit to Eq. S2. The extracted parameters are shown in Table S2. However, no clear $\mu_x$-induced spin-to-charge conversion signal is observed at these temperatures. As an example, the measurements at 300 K are shown in Figure S10.

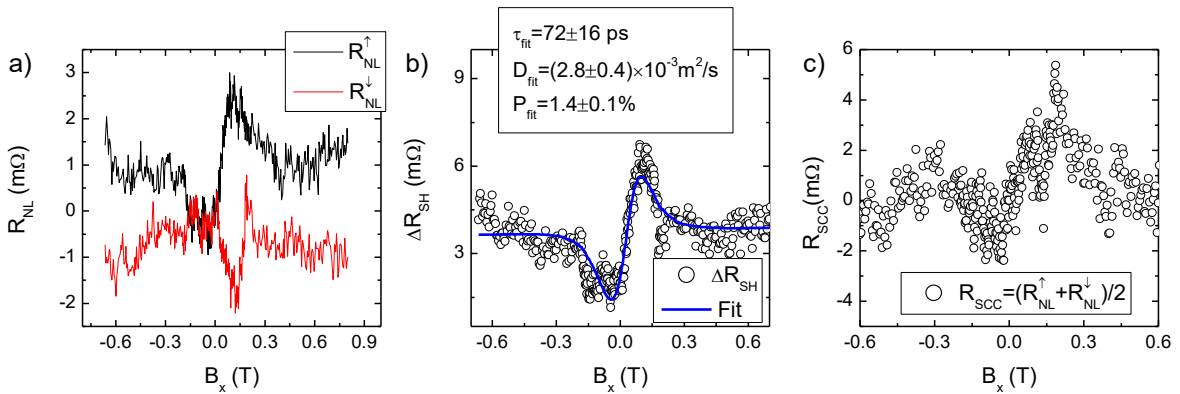

**Figure S10.** (a) Antisymmetric Hanle precession measurements for the two different orientations of the FM injector for sample B at 300 K. (b) $\Delta R_{SH}$ obtained as defined above by subtracting the components shown in (a) and its fit to Eq. S2 with the extracted parameters. (c) $R_{SCC}$. In this case $\Delta R_{SCC}$ is hard to extract because it is highly sensitive to the $B$ field range selected for the extraction. However, it can be estimated to be between 3 and 5 m$\Omega$.

## S6. Extracted parameters for sample A and B

**Table S1.** Extracted parameters for sample A. $D_s^{gr}$, $\tau_s^{gr}$, and $P_{sym}$ are extracted from fits to the symmetric Hanle precession data, $D_s^{eff}$, $\tau_s^{eff}$, and $P_{anti}$ are extracted from fits to the antisymmetric Hanle precession data, and $\Delta R_{SCC}$ is extracted from fits to the S-shaped background of the antisymmetric Hanle precession data. $\gamma$ and $\alpha_{RE}$ are obtained using $\Delta R_{SCC}$ and Eqs. S3-S4 with the spin transport parameters extracted from the antisymmetric Hanle precession data. The uncertainty ranges here are obtained considering only the uncertainty in $\Delta R_{SCC}$.

| Sample A | 10 K | 100 K | 200 K | 300 K |
|---|---|---|---|---|
| $R_{sq}^{gr}(\Omega)$ | 543 | 454 | 347 | 295 |
| $R_{sq}^{TMD/gr}(\Omega)$ | 497 | 430 | 335 | 288 |
| $D_s^{gr}(10^{-3}\,\text{m}^2/\text{s})$ | $5.7 \pm 0.5$ | $8.8 \pm 0.5$ | $8.8 \pm 0.5$ | $20 \pm 4$ |
| $\tau_s^{gr}$ (ps) | $300 \pm 20$ | $221 \pm 8$ | $221 \pm 8$ | $152 \pm 6$ |
| $P_{sym}(\%)$ | $1.4 \pm 0.1$ | $1.35 \pm 0.06$ | $1.35 \pm 0.06$ | $2.3 \pm 0.1$ |
| $P_{corr}(\%)$ | $2.9 \pm 0.2$ | $2.9 \pm 0.1$ | $2.9 \pm 0.1$ | $4.8 \pm 0.2$ |
| $D_s^{eff}(10^{-3}\,\text{m}^2/\text{s})$ | $1.2 \pm 0.1$ | | | |
| $\tau_s^{eff}$ (ps) | $72 \pm 7$ | | | |
| $P_{anti}(\%)$ | $3.6 \pm 0.2$ | | | |
| $\theta_{SH}^{TMD/gr}(\%)$ | $4.5 \pm 0.6$ | | | |
| $\sigma_{SH}^{Gr}(\Omega^{-1})$ | $(9 \pm 1) \times 10^{-5}$ | | | |
| $\Delta R_{SCC}(\text{m}\Omega)$ | $11.5 \pm 0.4$ | $12 \pm 0.5$ | $6.7 \pm 0.2$ | $7.4 \pm 0.5$ |
| $\gamma$(A/(Jm)) | $(2.9 \pm 0.1) \times 10^{20}$ | $(1.63 \pm 0.07) \times 10^{21}$ | $(3.02 \pm 0.09) \times 10^{21}$ | $(3.3 \pm 0.2) \times 10^{21}$ |
| $\alpha_{RE}(\%)$ | $0.85 \pm 0.03$ | $2.7 \pm 0.1$ | $3.32 \pm 0.1$ | $3.0 \pm 0.2$ |
| $\theta_{SH}^{MoS_2}(\%)$ | $3.3 \pm 0.1$ | $4.1 \pm 0.2$ | $2.98 \pm 0.09$ | $2.2 \pm 0.1$ |

**Table S2.** Extracted parameters for sample B. The "Left" and "Right" labels refer to the data obtained with the spin injector placed at the left and right sides of the TMD-covered region, respectively.

| Sample B | 10 K | 100 K | 200 K | 300 K |
|---|---|---|---|---|
| $R_{sq}^{gr}(\Omega)$ | | | | |
| $R_{sq}^{TMD/gr}(\Omega)$ | 282 | 266 | 231 | 215 |
| $D_s^{gr}(10^{-3}\,\text{m}^2/\text{s})$ | $7.0 \pm 0.3$ | $11 \pm 1$ | $15 \pm 2$ | $16 \pm 3$ |
| $\tau_s^{gr}$ (ps) | $252 \pm 8$ | $146 \pm 14$ | $139 \pm 2$ | $114 \pm 1$ |
| $P_{sym}(\%)$ | $2.4 \pm 0.1$ | $7.2 \pm 0.1$ | $7.7 \pm 0.1$ | $6.9 \pm 0.1$ |
| Right $D_s^{eff}(10^{-3}\,\text{m}^2/\text{s})$ | $1.5 \pm 0.1$ | $1.6 \pm 0.1$ | $1.4 \pm 0.1$ | $2.8 \pm 0.4$ |
| Right $\tau_s^{eff}$ (ps) | $70 \pm 9$ | $49 \pm 3$ | $44 \pm 3$ | $72 \pm 16$ |
| Right $P_{anti}(\%)$ | $3.4 \pm 0.3$ | $3.8 \pm 0.3$ | $3.1 \pm 0.3$ | $1.5 \pm 0.1$ |
| Right $\theta_{SH}^{TMD/gr}(\%)$ | $4.8 \pm 0.9$ | $2.0 \pm 0.5$ | $1.2 \pm 0.2$ | $0.33 \pm 0.04$ |
| Right $\sigma_{SH}(\Omega^{-1})$ | $(17 \pm 3) \times 10^{-5}$ | $(7.5 \pm 2) \times 10^{-5}$ | $(5.2 \pm 0.9) \times 10^{-5}$ | $(1.5 \pm 0.2) \times 10^{-5}$ |
| Left $D_s^{eff}(10^{-3}\,\text{m}^2/\text{s})$ | $2.1 \pm 0.1$ | $1.5 \pm 0.1$ | | |
| Left $\tau_s^{eff}$ (ps) | $73 \pm 6$ | $53 \pm 2$ | | |
| Left $P_{anti}(\%)$ | $1.9 \pm 0.1$ | $3.0 \pm 0.2$ | | |
| Left $\theta_{SH}^{TMD/gr}(\%)$ | $1.5 \pm 0.2$ | $1.3 \pm 0.2$ | | |
| Left $\sigma_{SH}^{Gr}(\Omega^{-1})$ | $(5.3 \pm 7) \times 10^{-5}$ | $(4.9 \pm 0.8) \times 10^{-5}$ | | |

# S7. Comparison between spin Hall effect in Pt and in TMD/graphene devices

## S7.1. TMD/graphene normalized conversion efficiency

As shown in Section S2, the signal induced by the spin Hall effect in the case of TMD/graphene heterostructures is determined as:

$$\Delta R_{SH} = \frac{\Delta V_{SH}}{I_c} = 2\theta_{SH}^{TMD/gr} R_{sq}^{TMD/gr} \frac{\bar{I}_s}{I_c}$$

We want to determine an efficiency factor which is a trans-resistance defined as follows: $R_{eff}^{TMD/gr} = \Delta V_{SH}/I_s$ where $I_s$ is the spin current entering the TMD/graphene heterostructure and $\Delta V_{SH} = (R_{NL}^{\uparrow} - R_{NL}^{\downarrow})I_c$ is the transverse voltage output measured as shown in Figure S3. Since the measured signal depends on $\bar{I}_s$, the average spin current at the region where the conversion takes place, we rewrite the equation above in the following way:

$$\Delta V_{SH} = \left(\frac{2\theta_{SH}^{TMD/gr} R_{sq}^{TMD/gr} \bar{I}_s}{I_s}\right) I_s = R_{eff}^{TMD/gr} I_s$$

The efficiency factor is defined as follows:

$$R_{eff}^{TMD/gr} = \frac{2\theta_{SH}^{TMD/gr} R_{sq}^{TMD/gr} \bar{I}_s}{I_s}$$

To obtain $R_{eff}^{TMD/gr}$ we need to determine the correction factor $\bar{I}_s/I_s$, which can be done by considering that the spin current entering the TMD-covered graphene region is: $I_s(x) = (P_i I_c e^{-x/\lambda_s^{eff}})/2$ and the average spin current at the region of interest reads:

$$\bar{I}_s = \frac{1}{W_{cr}} \int_L^{L+W_{cr}} I_s(x) dx = \frac{P_i I_c}{2 W_{cr}} \int_L^{L+W_{cr}} e^{-x/\lambda_s^{eff}} dx = \frac{P_i I_c \lambda_s^{eff}}{2 W_{cr}} \left(e^{-L/\lambda_s^{eff}} - e^{-(L+W_{cr})/\lambda_s^{eff}}\right)$$

Now we can obtain the $\bar{I}_s/I_s$ factor:

$$\frac{\bar{I}_s}{I_s} = \frac{\lambda_s^{eff}}{W_{cr}} \frac{e^{-L/\lambda_s^{eff}} - e^{-(L+W_{cr})/\lambda_s^{eff}}}{e^{-L/\lambda_s^{eff}}} = \frac{\lambda_s^{eff}}{W_{cr}}\left(1 - e^{-W_{cr}/\lambda_s^{eff}}\right)$$

which for sample A at 10 K is 0.3. Now we finally write the efficiency factor for the TMD/graphene case:

$$R_{eff}^{TMD/gr} = \frac{2\theta_{SH}^{TMD/gr} R_{sq}^{TMD/gr} \lambda_s^{eff}}{W_{cr}}\left(1 - e^{-W_{cr}/\lambda_s^{eff}}\right)$$

This expression gives a numerical value of $R_{eff}^{TMD/gr} \approx 13.4\ \Omega$ for sample A at 10 K. Note that, because we have measured with magnetic fields applied in the $x$ direction, there is still a correction factor required to account for the diffusive broadening of the spin precession, which we have estimated to be of about 0.62 using Eq. S2.

## S7.2. Pt/graphene normalized conversion efficiency

We would like to compare the expression obtained above for $R_{eff}^{TMD/gr}$ with a properly normalized efficiency for Pt/graphene devices in which the spin Hall signal is determined by [S9]:

$$\Delta R_{SH}^{Pt} = \frac{2\theta_{SH}^{Pt}\rho_{Pt}x_{Pt/gr}}{W_{Pt}}\frac{\bar{I}_s}{I_c}$$

Therefore, the efficiency factor can be obtained from this relation:

$$V_{SH}^{Pt} = \left(\frac{2\theta_{SH}^{Pt}\rho_{Pt}x_{Pt/gr}}{W_{Pt}}\frac{\bar{I}_s}{I_s}\right)I_s = R_{eff}^{Pt/gr}I_s$$

where, $I_s$ is the spin current entering the Pt layer, $\theta_{SH}^{Pt}$ the spin Hall angle, $\rho_{Pt}$ the resistivity, and $W_{Pt}$ the width of the Pt wire. $x_{Pt/gr}$ is a parameter that accounts for the parallel conduction channel provided by the graphene that can reduce the measured voltage. In the case of graphene, however, because its sheet resistance is higher than Pt, $x_{Pt/gr} \approx 1$. The $\bar{I}_s/I_s$ factor is [S9, S10]:

$$\frac{\bar{I}_s}{I_s} = \frac{\lambda_{Pt}}{t_{Pt}}\frac{1-e^{-t_{Pt}/\lambda_{Pt}}}{1+e^{-t_{Pt}/\lambda_{Pt}}}$$

where $t_{Pt}$ is the thickness of the Pt wire and $\lambda_{Pt}$ its spin diffusion length. Now we can write the efficiency for Pt as:

$$R_{eff}^{Pt/gr} = \frac{2\theta_{SH}^{Pt}\rho_{Pt}x_{Pt/gr}}{W_{Pt}}\frac{\lambda_{Pt}}{t_{Pt}}\frac{1-e^{-t_{Pt}/\lambda_{Pt}}}{1+e^{-t_{Pt}/\lambda_{Pt}}}$$

We estimate $R_{eff}^{Pt/gr} \approx 0.27\ \Omega$ using the following parameters from Ref. [S9]: $\theta_{SH}^{Pt} = 0.23$, $\rho_{Pt} = 134\ \mu\Omega$cm, $\lambda_{Pt} = 2$ nm, $W_{Pt} = 200$ nm, $t_{Pt} = 21$ nm.

The conversion efficiencies can now be directly compared by looking at the $R_{eff}^{Pt/gr}$ and $R_{eff}^{TMD/gr}$ values. Since $R_{eff}^{TMD/gr}$ is $\sim 50 \times R_{eff}^{Pt/gr}$, we conclude that the TMD-covered graphene is 50 times more efficient than highly resistive Pt for spin detection despite the fact that the spin Hall angle in Pt is an order of magnitude higher than in TMD-covered graphene. Note that, accounting for the spin precession used in our measurement leads to, $R_{eff}^{TMD/gr}/R_{eff}^{Pt/gr} = 31$, still a large difference.

## S8. Theoretical calculation of spin-to-charge conversion in TMD/graphene heterostructures

To obtain a theoretical estimate to the magnitude of SHE and REE signals in our devices, we use a simplified model that captures the physics of a monolayer graphene/TMD heterostructure by using the following Hamiltonian [S11, S12]:

$$H = H_0 + H_I + H_R$$

where

$$H_0 = \hbar v_f(\sigma_x k_x + \tau_z \sigma_y k_y) + \Delta\sigma_z$$

$$H_I = (\lambda_I \sigma_z + \lambda_{VZ})\tau_z s_z$$
$$H_R = \lambda_R(s_y \sigma_x k_x - \tau_z s_x \sigma_y)$$

with $\sigma_\alpha$, $\tau_\alpha$ and $s_\alpha$ the Pauli matrices on the $\alpha$ direction acting on the pseudospin, valley and spin degrees of freedom, respectively. $H_0$ represents the orbital part, which is described by the Dirac's Hamiltonian with Fermi velocity $v_f$ and a staggered potential with strength $\Delta$, the latter appearing because of the broken sub-lattice symmetry of the TMD. $H_I$ represents the intrinsic spin-orbit coupling of the heterostructure, modelled with a Kane-Mele term with strength $\lambda_I$ and a valley-Zeeman coupling characterized by the parameter $\lambda_{VZ}$, both appearing because of the honeycomb structure and the broken sub-lattice symmetry. This term is responsible for the SHE, as we will show ahead. Finally, $H_R$ represents the Rashba spin-orbit coupling arising from the interface between graphene and the TMD, with strength $\lambda_R$. This last term is the source of the REE. In this model we are excluding the so-called pseudospin inversion asymmetry terms [S12], because they do not contribute to any of the effects of interest at low energy, and we use the tight-binding parameters from Ref. [S12].

### S8.1. The spin Hall effect

Despite the presence of the Rashba spin-orbit coupling, the origin of the spin Hall effect was traced back to the intrinsic spin-orbit coupling [S11], described in these heterostructures by a valley-Zeeman and Kane-Mele spin-orbit coupling, which interacts with the staggered potential for producing a net berry phase and, therefore, a finite SHE. For simplicity, we can neglect the Rashba spin-orbit coupling, and consider the spin-up and spin-down bands as independent. Moreover, if we also neglect intervalley scattering, the Dirac's cone can also be considered as independent, leading then to 8 almost identical bands described by the dispersion relation

$$\varepsilon(\mathbf{k}, \tau, s) = s\lambda_{VZ} \pm \sqrt{\hbar(v_F k)^2 + \Delta_{s,\tau}^2},$$

where $\Delta_{s,\tau} = \Delta - \tau s \lambda_I$. Each of these bands will contribute to the total spin Hall conductivity, and the calculation can be done following Refs. [S13, S14], leading to the following result:

$$\sigma_{SH}(\varepsilon_F) = \frac{e^2}{2h}\sum_{s,\tau=\pm 1} \tau \left[ \Delta_{s,\tau} \frac{\Theta(|\Delta_{s,\tau}|-|\varepsilon_F-\tau s\lambda|)}{\sqrt{\Delta_{s,\tau}^2+(\varepsilon_F-\tau s\lambda)^2}} + \Theta(|\varepsilon_F - \tau s\lambda| - \Delta_{s,\tau}) \right] \quad (S5)$$

where $\Theta$ is the Heaviside function. The comparison between this model and the curve simulated by solving numerically the Kubo formula for graphene/MoS$_2$ heterostructure is presented in Figure S11, where the small difference arises from the difference between the broadening functions used in the model (Lorentzian) and in the simulation (Lorentzian-like) [S11]. Figure S12a shows the spin Hall conductivity over a range of Fermi energies for three different temperatures. At $T = 0$ one can see that the maximum of the spin Hall conductivity is achieved at two points at around $\pm 1$ eV, changing sign when crossing the charge neutrality point (CNP). Therefore, a negative spin Hall angle (as found in the main text) would be an indication of the Fermi level being at the valence band of graphene but close to the CNP. In addition, the spin Hall conductivity maximum is highly susceptible to temperature effects, where we see a linear variation with $\beta = 1/T$ at low and high temperature (Figure S12b). In this sense, the maximum spin Hall conductivity achieved in this system at low temperature is $0.2e^2/h$. This value is also suppressed by intervalley scattering [S15], so it is difficult to establish an exact value, but assuming weak intervalley scattering and using the square resistance ($R_{sq}^{TMD/gr}$) extracted from sample A, one obtains a conversion efficiency in the range of $\theta_{SH}^{TMD/gr} = 1 - 10\%$, which is consistent with the experimental observation.

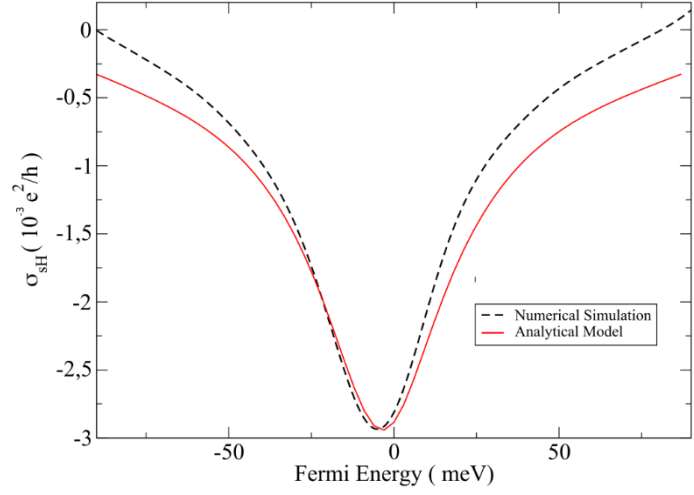

**Figure S11.** Comparison between the simulation of the spin Hall effect for graphene/MoS$_2$ heterostructure and the model based on individual spin bands. We have introduced a gaussian broadening of 20 meV to match the intrinsic broadening of the Kernel Polynomial Method (KPM) simulations performed at Ref. [S11]. The small difference is due to the broadening which is not energy independent in the KPM method.

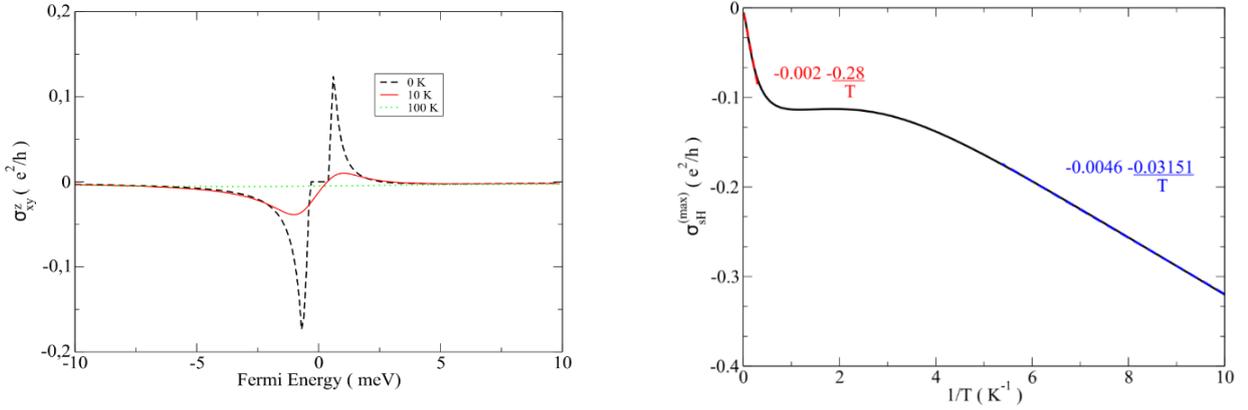

**Figure S12.** (a) Spin Hall conductivity as a function of Fermi energy for three different temperatures and (b) the evolution of the maximum of the spin Hall conductivity computed as a function of temperature, both computed using Eq. S5. Both results are obtained at zero gaussian broadening.

### S8.2. The Rashba-Edelstein effect and its reciprocal

The REE is a phenomenon where the combined effect of a momentum dependent spin texture, and an external electric field $\boldsymbol{E}_0$ produces a nonequilibrium spin density $\boldsymbol{S}^{(neq)}$. It is usually described by the electrical spin susceptibility $\chi_{RE}$, which under the linear response limit allow for writing the following constitute relation

$$\boldsymbol{S}^{(neq)} = \chi_{RE}(\boldsymbol{E}_0 \times \hat{k})\widehat{\boldsymbol{S}}^{(neq)}.$$

The polarization of the spin density is perpendicular to both the external electric field and the out-of-plane direction and is defined in units of a charge density. The REE is a dissipative effect and, as shown by Offidani et al. [S16], for pure Rashba-like systems this implies that the spin susceptibility is proportional to the DC conductivity ($\sigma_{DC}$)

$$\frac{\chi_{RE}}{\sigma_{DC}} = \frac{1}{v_{RE}},$$

where $v_{RE}$ is a parameter with units of velocity characterizing the spin-to-charge conversion efficiency. There are different estimations for this parameter. We have, for instance the result of Dyrdal [S13]:

$$\frac{v_F}{v_{RE}} = \frac{1}{2}\frac{2+x}{1+x}\Theta(2-x) + \frac{x}{x^2-1}\Theta(x-2), \text{ with } x \equiv \varepsilon_F/\lambda_R,$$

the result of Offidani et al. [S16]:

$$\frac{2v_F}{v_{RE}} = \Theta(2-x) + 2xg(D_c)\Theta(x-2), \text{ with } x \equiv \varepsilon_F/\lambda_R,$$

where $g(D_c)$ is a dimensionless parameter that depends of the disorder, and the result of Garcia et al. [S11], which takes into account all the spin-orbit parameters leading to a higher spin density for high energies but to a small $1/v_{RE}$ due to the broadening effects present in the system, which will increase $\sigma_{DC}$ close to the CNP.

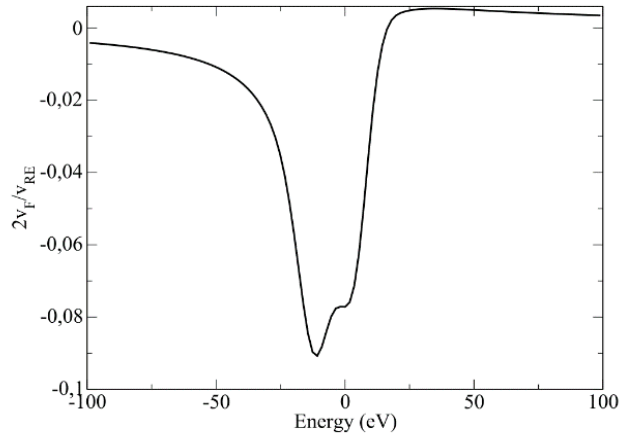

**Figure S14.** Figure of merit $2v_F/v_{RE}$ of the Rashba Edelstein effect for graphene/MoS$_2$ heterostructure considering a broadening of 5 meV.

Using $2v_F/v_{RE}$ as a figure of merit, it is possible to extract a conversion efficiency $\alpha_{RE}$ by computing the transformation from spin density to spin current using Fick's law [S1]:

$$\alpha_{RE} = \frac{J_s^y}{J_x} = \frac{D_s}{\lambda_s v_{RE}}$$

where $D_s$ and $\lambda_s$ are the spin diffusion constant and the spin relaxation length of graphene. One should keep in mind that in this process there is no electric field affecting the spin current; therefore, the charge current density is obtained solely from the spin density. Using $D_s$ and $\lambda_s$ obtained from sample A, one obtains a conversion efficiency $\alpha_{RE}$ ~0.1% in the best-case scenario, decreasing down to ~0.01% at higher temperatures. One should also keep in mind that there is an additional suppression term which appears due to the averaging over the channel width, which will further suppress this value.

The REE conversion efficiency $\alpha_{RE}$ changes sign following the same trend as the spin Hall angle $\theta_{SH}^{TMD/gr}$, because both effects originate from the proximity-induced spin-orbit coupling of graphene. Therefore, both $\alpha_{RE}$ and $\theta_{SH}^{TMD/gr}$ have the same sign. This is not observed in our experiment, ruling out the REE as the origin of the in-plane spin-to-charge conversion signal.

## S9. Scanning electron and atomic force microscopy imaging of sample A

After the electrical characterization we imaged our device with scanning electron microscopy (SEM), see Figure S15. We extracted the length and width of the graphene channels as well as the width of the ferromagnetic electrodes and their distances from these measurements. As the samples were in contact with air after the electrical characterization in vacuum, the Co electrodes are oxidized in the images.

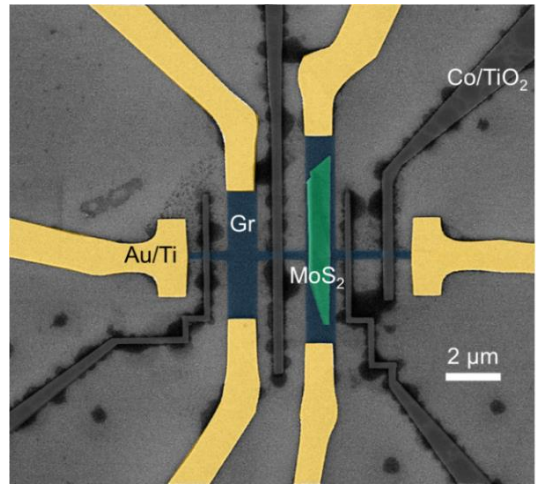

**Figure S15.** False-colored scanning electron microscopy image of Sample A. The width of the horizontal graphene (in blue) channel is 350 nm. The width of the $MoS_2$ flake (in green) and the two vertical graphene channels are 0.9 µm and 1.2 µm respectively.

After the electrical characterization and SEM imaging we measured the topography of device with an atomic force microscope (Agilent 5500) in tapping mode (Figure S16). The main result of this measurement is the height profile of the graphene/$MoS_2$ stack, as we extracted all lateral distances from the SEM image. The thickness of the graphene flake was determined to be roughly 5 nm. The $MoS_2$ flake shows a ramp-like shape on top and is between 35 and 75 nm thick.

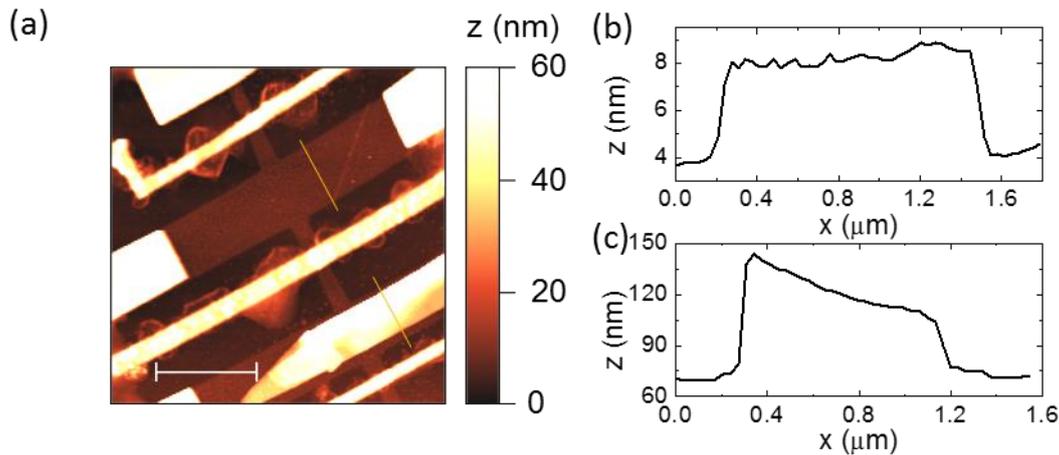

**Figure S16.** Atomic force microscopy characterization of Sample A. (a) Area scan showing the topography of the device. The scale bar is 2 µm long. (b) Line profile taken along the marked line in (a) across the graphene flake, where the thickness of the graphene flake is extracted to be roughly 5 nm, equivalent to around ten layers [S18]. (c) Line profile taken along the marked line in (a) across the $MoS_2$ flake, where the thickness of the $MoS_2$ flake is between 35 and 75 nm.